\documentclass[aps,nofootinbib,amsmath,prd,onecolumn,notitlepage,showpacs,superscriptaddress,groupedaddress,11pt]{revtex4-1}

\usepackage[american]{babel}
\usepackage{amsfonts}
\usepackage{amsmath}
\usepackage{amssymb}
\usepackage{slashed}
\usepackage{xcolor}
\usepackage{bm}
\usepackage{float}
\usepackage[utf8]{inputenc}
\usepackage{nicefrac}
\usepackage{hyperref}
\usepackage{url} 

\usepackage{mathtools}

\usepackage{graphicx,txfonts}

\definecolor{darkblue}{rgb}{0.1,0.1,0.7}
\hypersetup{colorlinks,
           linkcolor={darkblue},
           citecolor={darkblue},
           urlcolor={darkblue}
}

\newcommand{\lvec}[2]{\raise #1\hbox{$^\leftarrow$} \hspace{-9pt} #2}
\newcommand{\rvec}[2]{\raise #1\hbox{$^\rightarrow$} \hspace{-9pt} #2}
\newcommand{\lrvec}[2]{\raise #1\hbox{$^\leftrightarrow$} \hspace{-9pt} #2}

\allowdisplaybreaks

\usepackage[normalem]{ulem}

\newcommand{\mycomment}[1]{}

\allowdisplaybreaks

\begin{document}
%%%%%%%%%%%%%%%%%%%%%%%%%%%%%%%%%%%%%%%%%%%%%%%%%%%

\title{Auxiliary-Field Formalism for Higher-Derivative Boundary CFTs}

\author{Gregorio Paci}
\email{gregoriopaci@icloud.com}
\affiliation{
Institut Denis Poisson UMR 7013, Universit\'e de Tours, Parc de Grandmont, 37200 Tours, France}

\author{Sergey N. Solodukhin}
\email{sergey.solodukhin@univ-tours.fr}
\affiliation{
Institut Denis Poisson UMR 7013, Universit\'e de Tours, Parc de Grandmont, 37200 Tours, France}
\affiliation{  Institute for Theoretical and Mathematical Physics, 
Lomonosov Moscow State University, 119991 Moscow, Russia}

%%%%%%%%%%%%%%%
\begin{abstract}
%%%%%%%%%%%%%%%
%
We study the conformal field theory defined by the fourth-order operator  on four-dimensional manifolds with boundaries, reformulating it through an auxiliary field so that the dynamics become second order. Within this framework, we compute the heat kernel of $\Box^2$ in flat space exactly, together with the associated Seeley-DeWitt coefficients for a broad class of non-standard boundary conditions. On curved backgrounds, we further construct the Weyl-invariant completion of the auxiliary field action with boundary terms and identify the corresponding conformal boundary conditions. Finally, we compute the boundary charges in the trace anomaly from the displacement operator correlators. 
\end{abstract}

\maketitle

%\tableofcontents

%%%%%%%%%%%%%%%%%%%%%%%%%%%%%%%%%%%%%%%%%%%%%%%%%%%%%%%%%%
\section{Introduction}\label{sect:intro}
%\setcounter{equation}0
%%%%%%%%%%%%%%%%%%%%%%%%%%%%%%%%%%%%%%%%%%%%%%%%%%%%%%%%%%
Free conformal field theories (CFTs) with higher-derivative kinetic operators provide a controlled yet nontrivial setting in which to explore non-unitary realizations of conformal symmetry, the structure of trace anomalies, and the role of boundaries beyond second order. In four dimensions, the simplest example is the free scalar theory governed by the operator $\Box^2\phi$. This theory has a long history, with early applications in elasticity theory \cite{Landau:1986aog}, while its CFT aspects---together with those of their close relatives $\Box^k$---have been analyzed in flat space in \cite{Brust:2016gjy}, and in the presence of boundaries in \cite{Chalabi:2022qit} in dimension higher than four. A related interesting analysis has been performed in \cite{Gaikwad:2023gef}, where a boundary Liouville CFT has been considered. As a CFT, $\Box^2\phi$ in $d=4$ it is somewhat special. Conformal symmetry fixes the scalar 2-point function in general dimension to scale as
\begin{align}
\langle \phi(x)   \phi(0) \rangle
\sim
\frac{1}{|x|^{d-4}}
\, ,
\end{align}
which degenerates to a constant in $d=4$, ultimately leading to a vanishing stress tensor \cite{Brust:2016gjy} and prevents the standard conformal data from being extracted in the usual way. To discuss the conformal anomaly in this setting, it is therefore natural adopt a perspective close to logarithmic CFTs (see  \cite{Hogervorst:2016itc} for a review) and to consider logarithmic 2-point functions. As a matter of fact, the logarithmic structure will appear quite naturally in our framework employing auxiliary scalar fields since the primary $\phi$ and the auxiliary field necessary to lower the number of derivatives are closely related to the logarithmic multiplets typical of logarithmic CFTs \cite{Bergshoeff:2011xy}. However, while logarithmic $2$-point functions will appear, we will not insist on interpreting our theory as a full-fledged logarithmic CFT.
%Interestingly, in logCFTs we have a scale entering in the theory to render logarithmic arguments dimensionless. However, this scale carries no physical meaning, amounting just to a reparametrization of the logarithmic multiplets, see \cite{Hogervorst:2016itc}.

Boundaries further enrich the structure of quantum and conformal field theories, and find physical applications in areas ranging from condensed matter to string theory, see \cite{Cardy:1984bb,McAvity:1992fq,McAvity:1995zd}. 
For example, the presence of a boundary modifies the Ward identities, alters the operator content, and introduces new anomaly contributions localized at the boundary and the related ``conformal charges'' \cite{Solodukhin:2015eca}, which  are of interest for entanglement entropy and renormalization group flows. %For higher-derivative theories, boundaries are even more delicate because the variational principle and conformal symmetry severely constrain admissible boundary conditions.

A well known fact to bear in mind is that Weyl invariance in curved space implies conformal invariance, in the sense that the flat space limit of a Weyl invariant action yield a conformal field theory \cite{Osborn_lectures}.
The curved-space uplift of the flat operator $\Box^2$ produces the Paneitz operator $\Delta_4$, also known as Fradkin-Tseytlin-Paneitz-Riegert operator \cite{Fradkin:1982xc,Riegert:1984kt,paneitz}. This is the first nontrivial higher-order operator in the Graham--Jenne--Mason--Sparling (GJMS) hierarchy \cite{GJMS,gover-hirachi}, whose structural properties allows to construct anomaly-induced effective actions in arbitrary even dimension \cite{Paci:2024xxz}. 
Conformal boundary operators associated with $\Delta_4$ were analyzed in \cite{Case:Paneitz_Bop}, and reduce in flat space to the classification of boundary operators obtained in \cite{Chalabi:2022qit}. These geometric insights will be important for our purposes.

In this work we revisit boundary higher-derivative CFTs from the perspective of the auxiliary-field formalism, proposed in \cite{Gibbons:2019lmj} for higher derivatives quantum field theories. Introducing an auxiliary scalar to reduce a fourth-order action to a second-order system, greatly simplify explicit computations---for example of the stress tensor and heat-kernel coefficients. However, this reformulation may raise the issue of to what extent the auxiliary formulation is fully equivalent to the original four-order theory. Standard arguments usually establish such an equivalence on-shell or at the level of the equation of motion, whereas many objects of interest in QFT---such as correlators and Ward identities---are normally computed off-shell. For instance, to meaningfully discuss the conformal anomalies one should first extend Weyl invariance to hold also off-shell in the auxiliary formulation. We anticipate that to achieve this result we will show that the Weyl invariant action is not quadratic in the auxiliary field.
Moreover, even if the auxiliary action is quadratic, for Gaussian integration to guarantee the equivalence of the correlators in the presence of a boundary, it is important to establish a precise mapping between the boundary conditions imposed in the four-order theory and in the auxiliary formulation.

Within this framework we develop a systematic treatment of higher-derivative CFTs on manifolds with boundary, centered on the interplay between auxiliary fields, conformal symmetry, and boundary data. In flat space, the auxiliary field formulation enables us to compute the heat kernel and the Seeley–DeWitt coefficients for quite general boundary conditions, including non-conformal ones.
By coupling the auxiliary action to a background metric and assigning a suitable Weyl transformation to the auxiliary field, we construct a curved-space Weyl-invariant action which is generically non-quadratic in the auxiliary field but whose flat-space limit reproduces the auxiliary formulation of the $\Box^2\phi$ theory. This construction also naturally identifies the associated conformal boundary conditions.
Finally, by analyzing displacement-operator correlators---which in this setup involve both $\phi$ and the auxiliary field and their correlators---we extract the corresponding boundary anomaly coefficients and show that they noetheless coincide with those obtained in the original fourth-order theory under the corresponding boundary conditions.

The paper is organized as follows.
Sect.~\ref{sect:anomaly_flat_space} is devoted to the flat-space analysis. In subsect.~\ref{subsect:reduction} we introduce the auxiliary-field action, obtain a general class of boundary conditions, and discuss how they map to those of the original biharmonic theory. We also diagonalize the kinetic term of the auxiliary action by performing a rotation in field space, and analyze how the boundary conditions transform under this rotation.
Exploiting this diagonalization, in subsect.~\ref{subsect:heat_kernel_flat_space} we compute explicitly the full heat kernel and the resulting effective action for $\Box^2$ under a variety of boundary conditions. 

Sect.~\ref{sect:anomaly_curved_space} turns to the properties of the auxiliary action under Weyl transformations. In subsect.~\ref{subsect:weyl_invariance_auxiliary_field} we determine the Weyl transformation of the auxiliary field and construct the corresponding Weyl-invariant extension of the auxiliary action. Conformal boundary conditions are also analyzed in this context. In subsect.~\ref{subsect:boundary_charges_auxiliary} we compute the energy-momentum tensor and extract the boundary central charges from 2- and 3-point functions of the displacement operator. 

Finally, in Sect.~\ref{sect:conclusions_outlook} we present our conclusions and outlook. Many additional technical results used througout the paper are reported in the Appendices. The Appendix~\ref{app:green_formula_with_boundaries} discuss the Green's formula on manifolds with boundaries, showing the boundary conditions one should impose on the Green function depending on those imposed on the field. In Appendix~\ref{appt:fourier_Integrals} we report the computations of the 2-point functions of $\phi$ and the auxiliary field, we show in particular how the logarithmic term appear. Appendix~\ref{app:emt_auxiliary_fields} discusses the metric variation of the axuliary action under the simplest pair of conformal boundary conditions, and Appendix~\ref{app:2_pt_function_displacement} gives a brief summary of the computation of the displacement correlators.

%%%%%%%%%%%%%%%%%%%%%%%%%%%%%%%%%%%%%%%%%%%%%%%%%%%%%%%%%%
\section{Auxiliary Fields in Higher-Derivative Boundary CFTs: Flat Space}\label{sect:anomaly_flat_space}
%\setcounter{equation}0
%%%%%%%%%%%%%%%%%%%%%%%%%%%%%%%%%%%%%%%%%%%%%%%%%%%%%%%%%%
We start by considering the bi-harmonic operator on scalars in flat space $\mathbb{R}^d$ with a flat boundary $\partial M=\{z=0\}$, i.e, $M=\mathbb{R}^{d-1} \times \mathbb{R}^+$. We will denote with $n$ the outward normal vector, while $\partial_n=\partial_z$ is the normal derivative. The action reads
\begin{align}\label{eq:bi_harmonic_flat_space_action}
\frac{1}{2}
\int_M d^d x
\phi
\Box^2
\phi
\, .
\end{align}
For the operator $\Box^2$ to be symmetric with respect to the standard scalar product, we want the following symplectic form to vanish 
\begin{align}\label{eq:bi_harmonic_flat_space_boundary_terms}
\omega_{\Box^2}
=
\int_M d^d x
\left(
\phi_1
\Box^2
\phi_2
-
\phi_2\Box^2\phi_1
\right)
=
\int_{\partial M} d^{d-1} x
\left(
\phi_1 \partial_n \Box\phi_2
-
\partial_n \phi_1  \Box\phi_2
-
\phi_2 \partial_n \Box \phi_1
+
\partial_n \phi_2  \Box\phi_1
\right)
\, .
\end{align}
We now list all the couples of boundary conditions which kill the boundary terms, thus leading to a symmetric operator. For the moment we ignore conformal invariance, which will be considered later. Our main concern here is to outline the strategy we will use to study boundary conditions throughout the paper.

Let us consider the boundary operators
\begin{align}\label{eq:BOs_flat_space_biharmonic}
\Phi^0 &= \phi|_{\partial M}\, ,    &     \Phi^1 &= \partial_n \phi|_{\partial M} \,,   \\
\Phi^2 &= \Box\phi|_{\partial M} \, ,     &     \Phi^3 &= \partial_n \Box\phi|_{\partial M} \,,  \nonumber
\end{align}
and define the vectors $\Psi$, $\Phi$ and the matrix $J$ as
\begin{align}
\Psi
=
\begin{bmatrix} \Phi^0  \\ \Phi^2 \end{bmatrix}
\, ,
\quad
\Phi
=
\partial_n \Psi
=
\begin{bmatrix} \Phi^1  \\ \Phi^3 \end{bmatrix}
\, ,
\quad
J
=
\begin{bmatrix} 0 & 1  \\ 1 & 0 \end{bmatrix}
\, .
\end{align}
Then, Eq.~\eqref{eq:bi_harmonic_flat_space_boundary_terms} can be compactly written as
\begin{align}\label{eq:omega_fourth_order_formulation}
\omega_{\Box^2}
=
\int_{\partial M} d^{d-1} x
\left(
\Psi^T_1 J \Phi_2
-
\Psi^T_2 J \Phi_1
\right)
=
\langle \Psi_1, J \Phi_2 \rangle
-
\langle \Psi_2, J \Phi_1 \rangle
\, ,
\end{align}
where $\langle \, ,  \rangle$ denotes the standard scalar product over $\partial M$. The scalar product in the bulk will be denoted with round brackets.
Other choices for $\Psi$ and $\Phi$ are possible but physically equivalent.
With this choice and for boundary conditions of the form 
\begin{align}\label{eq:generic_BC_fourth}
\Phi=
S
\Psi 
\, ,
\end{align}
the requirement $\omega_{\Box^2}=0$ leads to
\begin{align}\label{eq:J_symmetric_S}
(JS)^T=JS
\quad \Rightarrow \quad
S^T=JSJ^{-1}
\quad \Rightarrow \quad
S
=
\begin{bmatrix} s & s_{12}  \\ s_{21} & s \end{bmatrix}
\, .
\end{align}
We will call a matrix satisfying such a condition a $J$-symmetric matrix.
We stress that \eqref{eq:J_symmetric_S} is only a requirement on the ``form'' of $S$, meaning that in principle the entries might be formally infinite, zero or also differential operators. The important point is that they must have the correct mass dimension to match the number of derivatvives. %We will see an example when studying conformal boundary conditions. 

We also notice that in principle more general conditions setting $\omega_{\Box^2}=0$ are possible. We can rewrite \eqref{eq:omega_fourth_order_formulation} as
\begin{align}
\omega_{\Box^2}
=
\langle X, \Omega Y \rangle
%=
%X^T \Omega Y
%
\, ,
\qquad\qquad
X
=
\begin{bmatrix}  \Psi_1   \\  \Psi_2  \end{bmatrix}
\, ,
\quad
Y
=
\begin{bmatrix}  \Phi_1   \\  \Phi_2  \end{bmatrix}
\, ,
\quad
\Omega
=
\begin{bmatrix} 0 & J  \\ -J & 0 \end{bmatrix}
\, .
\end{align}
Imposing $\omega_{\Box^2}=0$, for $Y=QX$, would impose $(\Omega Q)^T=-\Omega Q$ or 
\begin{align}
Q^T=\Omega Q \Omega^{-1}
\qquad
\Rightarrow
\qquad
Q
=
\begin{bmatrix} S & A_1  \\ A_2 & S \end{bmatrix}
\, ,
\end{align}
where $S$ and $A_{1,2}$ are $J$-symmetric and $J$-antisymmetric, respectively. Then, we have the BCs
\begin{align}
(\Phi_1=S\Psi_1 + A_{1}\Psi_2  \, , \quad \Phi_2=A_{2}\Psi_1 + S\Psi_2)
\, ,
\end{align}
mixing $(1,2)$ components (observe that $(1,2)$ are subscripts). To avoid this, we are forced to choose $A_{1,2}=0$, effectively returning to the case \eqref{eq:generic_BC_fourth}.

Thus, we consider boundary conditions of the form \eqref{eq:generic_BC_fourth} which can be more explicitly written as
\begin{align}\label{eq:general_BC_biharmonic}
(\Phi^1=s\Phi^0 + s_{12}\Phi^2 \, , \quad \Phi^3=s_{21}\Phi^0 + s\Phi^2)
\, .
\end{align}
It is simple to verify explicitly that any choice of the $S$ entries in Eq.s~\eqref{eq:general_BC_biharmonic} leads to correct boundary conditions.
These boundary conditions require the entries of $S$ to have different mass dimension according the number of derivatives entering in $\Phi^i$. To avoid the presence of dimensionful parameters, as expected from regular conformal invariance, we should impose boundary conditions of the form $(\Phi^i=0, \Phi^j=0)$. After a quick inspection of the six possible couples, we arrive at all the possible scale invariant condtions
\begin{align}\label{eq:BCs_flat_space_biharmonic1}
(&\Phi^0=0, \Phi^1=0) \, , & (&\Phi^2=0, \Phi^3=0)\, , \\
 (&\Phi^0=0, \Phi^2=0)  \, , & (&\Phi^1=0, \Phi^3=0)   \label{eq:BCs_flat_space_biharmonic2}
\, .
\end{align}
Notice that $(\Phi^0=0, \Phi'^1=0) \equiv (\Phi^0=0, \Phi^1=0)$, and $(\Phi^2=0, \Phi'^3=0) \equiv (\Phi^2=0, \Phi^3=0)$. In terms of $S$, these correspond to the choices
\begin{align}\label{eq:choices_S_no_massive_parameters}
(&s=s_{12}=0, s_{21} \to \infty) \, , & (&s=s_{21}=0, s_{12} \to \infty)\, , \\
 (&s_{21}=s_{12}=0, s \to \infty)  \, , & (&s_{21}=s_{12}=s=0)  
\, ,
\end{align}
respectively. We also see that no consistent choice of $S$ can reproduce $(\Phi^0=0, \Phi^3=0)$ and $(\Phi^1=0, \Phi^2=0)$.\footnote{For instance, from the second of Eq.s~\eqref{eq:general_BC_biharmonic}, one could set $\Phi^3=0$ by taking $s_{21}=s=0$, but this also forces $\Phi^0$ to disappear from the first of Eq.s~\eqref{eq:general_BC_biharmonic}. More generally, to set $\Phi^3=0$, we might choose $\Phi^0=-\frac{s}{s_{21}}\Phi^2$. When substituted into the first equation, it yields $\Phi^0=\frac{s}{s^2+s_{12}s_{21}}\Phi^1 \overset{s_{12}\to \infty}{\to} 0$. This in turn implies $\Phi^2=0$ and thus $\Phi^1=0$ too. In other words, enforcing $(\Phi^0=0, \Phi^3=0)$ results in overconstraining the system, leading to $\Phi^{i}=0$ for all components. It is simple to verify that neither of these BCs set $\omega=0$.}

If we allow for the presence of dimensionful parameters, infinitely many boundary conditions are possible. We now consider some possibilities which will re-encounter later.
The Robin-like conditions are
\begin{align}
 (\Phi^3= s_{21}\Phi^0,    {\Phi}^1= s_{12}{\Phi}^2) \, ,
\qquad 
 (\Phi^1=s \Phi^0, \Phi^3=s \Phi^2)
\, ,
\end{align}
corresponding to $s=0$ and $s_{12}=s_{21}=0$ respectively.
Notice that in the last condition the form of $S$ correctly requires the same $s$  to appear. This is in fact necessary as
\begin{align}
\int_{\partial M} d^3 x
\left(
\phi_1 \partial_n \Box\phi_2
-
\partial_n \phi_1  \Box\phi_2
\right)
=
\int_{\partial M} d^3 x
\left(
\phi_1 s' \Box\phi_2
-
s\phi_1  \Box\phi_2
\right)
=
0
\qquad
\Rightarrow
\qquad
s=s'
\, .
\end{align}

To obtain boundary conditions mixing $\Phi^1$ with $\Phi^3$, and $\Phi^0$ with $\Phi_2$ we can consider the linear combinations 
\begin{align}\label{eq:general_BC_biharmonic_llinear_combinations}
(\Phi^1+ \alpha \Phi^3=(s+\alpha s_{21})\Phi^0 + (s_{12}+\alpha s)\Phi^2) \, , \qquad (\Phi^1- \alpha \Phi^3=(s-\alpha s_{21})\Phi^0 + (s_{12}-\alpha s)\Phi^2)
\, ,
\end{align}
where $\alpha$ is a dimensionful function. Imposing $s_{21}=0$ and that both left and right hand-sides of these equations vanish, we obtain
\begin{align}\label{eq:chiral_BC_orginal_formulation}
(\Phi^0=\alpha \Phi^2, \Phi^1=-\alpha(\Phi^3+ \beta \Phi^2))
\, ,
\end{align}
where $\beta=-\frac{s_{12}}{\alpha}$. It can be directly verified that this choice set $\omega_{\Box^2}=0$.
%While the first set of boundary conditions we listed are compatible with scale invariance, and in fact conformal invariant, the others are not as they require to have non zero scale in the theory. However, in logCFT we do have a mass scale at our disposal which is necessary to make the argument of the logarithm dimensionless.
Our next goal is to compute the heat kernel for this operator under different boundary conditions using an auxiliary field.

\subsection{Reduction of the Higher-Derivative Action}\label{subsect:reduction}
We use an auxiliary scalar field to construct a theory equivalent to Eq.~\eqref{eq:bi_harmonic_flat_space_action} but with only second order derivatives. In this context with equivalent we mean they are equal on-shell \cite{Gibbons:2019lmj}.

Under appropriate BCs, from Eq.~\eqref{eq:bi_harmonic_flat_space_action} we have the equations of motion
\begin{align}\label{eq:4_ord_BCs_compatible_and_eom}
\quad
\Box^2
\phi
=
0
\, .
\end{align}
This fourth order equation is equivalent to the second-order system
\begin{align}\label{eq:2_ord_BCs_compatible_eom}
\Box
\psi
=
0
\, ,
\quad
\Box
\phi
=
\psi
\, ,
\end{align}
where $\psi$ is an auxiliary scalar field of dimension 2. We now construct an action $S(\phi,\psi)$ yielding these equations also in the presence of boundaries. A minimal symmetric choice is
\begin{align}\label{eq:second_ord_action_psi}
S[\phi,\psi]
=
\tfrac{1}{2}
\int_M d^4 x
\left(
\phi
\Box
\psi
+
\psi
\Box
\phi
-
\psi^2
\right)
\equiv
\tfrac{1}{2}
\int_M d^4 x
\Psi^T
{\hat{D}}
\Psi
\, ,
\end{align} 
where we introduced
\begin{align}\label{eq:second_ord_operator_psi}
\Psi= \begin{bmatrix} \phi  \\ \psi \end{bmatrix}
\, ,
\quad
{\hat{D}}= \begin{bmatrix} 0 & \Box  \\ \Box & -1 \end{bmatrix}
\, .
\end{align}
To have a symmetric operator we must impose 
\begin{align}
\left(\Psi_2, {\hat{D}} \Psi_1\right) =  \left({\hat{D}}\Psi_2, \Psi_1\right)
\, ,
\end{align}
where $(, \,)$ is the scalar product over $M$.
Integrating by parts we get the boundary terms
\begin{align}\label{eq:boundary_terms}
\omega_{\hat{D}}
=
\int_{\partial M} d^3 x
\left(
\phi_1\partial_n \psi_2
-
\psi_2\partial_n \phi_1
+
\psi_1\partial_n \phi_2
-
\phi_2\partial_n \psi_1
\right)
=
\langle \Psi_1, J\Phi_2 \rangle
-
\langle \Psi_2, J\Phi_1 \rangle
\, ,
\end{align}
where $\Phi_i=\partial_n\Psi_i$. With the same reasoning of the previous case, the boundary conditions are of the form
\begin{align}\label{eq:BC_general_form_second_order_formulation}
(\partial_n\phi=s\phi + s_{12}\psi \, , \quad \partial_n\psi=s_{21}\phi + s\psi)|_{\partial_M}
\, ,
\end{align}
which are in one-to-one correspondence with Eq.~\eqref{eq:general_BC_biharmonic} once we go on-shell for $\psi$.
The boundary conditions independent of trhe entries of $S$, i.e., scale invariant, now read 
\begin{align}
&( \phi|_{\partial M}=0, \partial_n\phi|_{\partial M}=0)\, , & &(\psi|_{\partial M}=0, \partial_n\psi|_{\partial M}=0) \label{eq:bad_BC_flat_space_auxiliary_field}    \, , \\
 &(\phi|_{\partial M}=0, \psi|_{\partial M}=0)  \, , & &(\partial_n\phi|_{\partial M}=0, \partial_n\psi|_{\partial M}=0)\label{eq:good_BC_flat_space_auxiliary_field}
\end{align}
with the same parameters choice as in Eq.s~\eqref{eq:choices_S_no_massive_parameters}. Choosing $s=0$ and $s_{12}=s_{21}=0$ we have instead the Robin-like BCs
\begin{align}\label{eq:robin_like_BCs_second_order_formulation}
(\partial_n\psi= s_{21} \phi,    \partial_n\phi= s_{12} \psi)\, ,
\qquad 
(\partial_n\psi= s \psi,    \partial_n\phi= s \phi) 
\, ,
\end{align}
respectively. Taking linear combinations of Eq.s~\eqref{eq:BC_general_form_second_order_formulation}, we also obtain chiral BCs such as
\begin{align}\label{eq:off_diagonal_BC_flat_space_auxiliary_field1}
%( \phi|_{\partial M}= \alpha\psi|_{\partial M}, \,\, \partial_n \phi|_{\partial M}= -\alpha \partial_n \psi|_{\partial M})
%\qquad 
( \phi|_{\partial M}= \alpha\psi|_{\partial M}, \,\, (\partial_n + s) \phi|_{\partial M}= -\alpha (\partial_n + s) \psi|_{\partial M}) 
\,  ,
\end{align}
corresponding to Eq.~\eqref{eq:chiral_BC_orginal_formulation}.
%We stress again that on-shell all of these BCs are mapped into those we found for the fourth order formulation. For example,  $(\phi|_{\partial M}=0, \psi|_{\partial M}=0)$ implies $(\phi|_{\partial M}=0, \Box \phi|_{\partial M}=0)$, matching the original fourth-order formulation. 

This is not of course an exhaustive list of boundary conditions. Another type of conditions we will encounter in the following is
\begin{align}
( (\partial_n + S) \phi|_{\partial M}= \alpha (\partial_n + S) \psi|_{\partial M}, \,\, (\partial_n + S')\phi|_{\partial M}= -\alpha(\partial_n + S')\psi|_{\partial M}) \label{eq:off_diagonal_BC_flat_space_auxiliary_field3}%
\,  ,
\end{align}
which can be re-expressed as
\begin{align}
\partial_n \phi = -S^+\phi +\alpha S^- \psi \, , \qquad \partial_n \psi = \frac{1}{\alpha}S^-\phi - S^+ \psi \, , \qquad\qquad  S^{\pm}=\frac{S\pm S'}{2} \, .
\end{align}
Thus, identifying $s=-S^+$ e $s_{12}=\alpha S^-$ e $s_{21}=\frac{1}{\alpha} S^-$, they are actually also of the general form \eqref{eq:BC_general_form_second_order_formulation}.

A few remarks are in order. Although the boundary conditions in Eq.s~\eqref{eq:bad_BC_flat_space_auxiliary_field} formally match those in Eqs.~\eqref{eq:BCs_flat_space_biharmonic1} in the auxiliary field formalism they must be discarded. In fact, in the second-order formulation they would overconstrain the system. since we are simultaneously requiring two boundary conditions on a scalar satisfying a second order equation. In Appendix~\ref{app:green_formula_with_boundaries} we study the Green formula for general operators on manifolds with boundaries---which determines the admissible boundary data---and provide detailed specific for the operators analyzed in this work.
On the other hand, the boundary conditions given, for instance, in Eqs.~\eqref{eq:good_BC_flat_space_auxiliary_field}, \eqref{eq:robin_like_BCs_second_order_formulation}, \eqref{eq:off_diagonal_BC_flat_space_auxiliary_field1}, and \eqref{eq:off_diagonal_BC_flat_space_auxiliary_field3} are admissible. 

Importantly, we anticipate that we will see from the results of Sect.~\ref{sect:anomaly_curved_space} that the boundary conditions \eqref{eq:good_BC_flat_space_auxiliary_field} are not only scale but also conformal invariant, and the action Eq.~\eqref{eq:second_ord_action_psi} equipped with one of \eqref{eq:good_BC_flat_space_auxiliary_field} defines a proper boundary CFT.

\subsubsection{Field Space Rotation and Diagonalization of the Kinetic Terms}

To compute the heat kernel in the presence of boundaries it proves convenient to use a different field parametrization. We consider the field space rotation-like transformation
\begin{align}\label{eq:field_space_rotation}
\Psi=
R(\lambda,\theta)
\Psi'
\, ,
\quad
\text{with}
\quad
R(\lambda,\theta)
=
\begin{bmatrix}  \cos\theta & \lambda\sin\theta   \\ -\frac{1}{\lambda}\sin\theta & \cos\theta   \end{bmatrix}
\, ,
\quad
\Psi'= \begin{bmatrix} \phi'  \\ \psi' \end{bmatrix}
\, ,
\end{align}
where $\lambda$ a parameter with the dimensions of length squared necessary to combine the fields $\phi$ and $\psi$. The advantage of the reparametrization using the fields $\phi'$ and $\psi'$ is that the kinetic matrix becomes diagonal for $\theta=\pi/4$. Up to a global $1/2$ factor, with this choice the action become
\begin{align}\label{eq:rotated_theory_before_normalization_and_continuation}
S'
=
\int_M d^4 x
\left(
-\frac{1}{\lambda}\phi' \Box \phi'
-
\frac{1}{2\lambda^2} \phi'^2
+
\lambda \psi' \Box \psi'
-
\frac{1}{2} \psi'^2
+
\frac{1}{\lambda}\phi'\psi'
\right)
\, .
\end{align}
Similarly, the boundary term \eqref{eq:boundary_terms} rotates as
\begin{align}
\omega'_{\hat{D}}
=
\langle \Psi'_1, \tilde{J}\Phi'_2 \rangle
-
\langle \Psi'_2,  \tilde{J}\Phi'_1 \rangle
\, ,
\quad
\tilde{J}
=
\begin{bmatrix} -\frac{1}{\lambda} & 0  \\ 0 & \lambda \end{bmatrix}
\, ,
\end{align}
where $\tilde{J}=R^T(\lambda,\pi/4)JR(\lambda,\pi/4)$ and $\Phi'=\partial_n\Psi'$.
For BCs of the form $\Phi'=\tilde{S}_\lambda\Psi'$, $\omega'_{\hat{D}}=0$ requires $\tilde{S}^T=\tilde{J} \tilde{S}_\lambda \tilde{J}^{-1}$, or, more explicitly,
\begin{align}\label{eq:tildeS}
\tilde{S}_\lambda
=
\begin{bmatrix} \tilde{s}_1 & \tilde{s}  \\ -\frac{\tilde{s}}{\lambda^2} & \tilde{s}_2 \end{bmatrix}
\, .
\end{align}
Of course, by integrating Eq.~\eqref{eq:rotated_theory_before_normalization_and_continuation} by parts we obtain the same results, and appling the rotation \eqref{eq:field_space_rotation} to the BCs \eqref{eq:BC_general_form_second_order_formulation} we get
\begin{align}
\Phi'=R(\lambda,\pi/4)SR^{-1}(\lambda,\pi/4)\Psi'
\, ,
\qquad
RSR^{-1}
=
\frac{1}{2}
\begin{bmatrix} 2s -(\lambda s_2 + s_1/\lambda)& \,\,\,\,\, s_1-\lambda^2s_2  \\ -\frac{1}{\lambda^2}(s_1-\lambda^2s_2) &  \,\,\,\,\,2s +(\lambda s_2 + s_1/\lambda) \end{bmatrix}
\equiv
\tilde{S}_\lambda
\,
.
\end{align}
Thus, $RSR^{-1}$ has the same form of $\tilde{S}$. This ensures that boundary conditions of the rotated theory are in one-to-one correspondence with those before the rotation.

To have canonically normalized kinetic terms we now choose $\frac{1}{\lambda}=m^2$ and rescale the fields as $\phi''=m\phi'$ and $\psi''=\frac{1}{m}\psi'$. Finally, renaming back $\phi''\equiv\phi'$ and $\psi''\equiv\psi'$ to keep the notation simple, we obtain
\begin{align}\label{eq:equivalent_2_order_problem_matrix_form_Kappa}
\int_M d^4 x
\left(
-\phi' \Box \phi'
-
\frac{m^2}{2} \phi'^2
+
\psi' \Box \psi'
-
\frac{m^2}{2} \psi'^2
+
m^2\phi'\psi'
\right)
\equiv
\int_M d^4 x
\Psi'^T
\hat{\mathcal{D}}
\Psi'
\, ,
\end{align}
where now
\begin{align}\label{eq:D_rotated_ghost}
\hat{\mathcal{D}}
=
R(\theta)^{\rm T} \hat{D} R(\theta)
=
 \begin{bmatrix} -\Box - \frac{m^2}{2} & \frac{m^2}{2}  \\ \frac{m^2}{2} & \Box -\frac{m^2}{2} \end{bmatrix}
\, ,
\end{align}
which highlights the ghost nature of $\psi'$. To avoid the ghost, which would render the heat kernel ill-defined, we analitically continue $\psi' \to i \psi'$ \cite{Gibbons:1978ac,Mazur:1989by}, arriving at the operator
\begin{align}\label{eq:D_rotated_continued}
\mathcal{\hat{D}}= \begin{bmatrix} -\Box - \frac{m^2}{2} &  i \frac{m^2}{2}  \\  i \frac{m^2}{2} & -\Box + \frac{m^2}{2} \end{bmatrix}
=
\begin{bmatrix} -\Box  & 0   \\0& -\Box \end{bmatrix}
+
\frac{m^2}{2}
\begin{bmatrix} -1 &  i    \\  i   &  1\end{bmatrix}
\equiv
%\mathcal{\hat{D}}_0
-\hat{1}\Box
+
\mathcal{\hat{M}}
\, .
\end{align}
Notice that 
\begin{align}
\mathcal{\hat{M}}
=
-
\frac{m^2}{2}(\sigma_3 - i \sigma_1)
\, ,
\end{align}
where we used the Pauli matrices
\begin{align}\label{eq:pauli_matrices}
\sigma_1
=
\begin{bmatrix}  0 &  1   \\  1   &  0 \end{bmatrix}
\, ,
\quad
\sigma_2
=
\begin{bmatrix}  0 &  -i   \\  i   &  0 \end{bmatrix}
\, ,
\quad
\sigma_3
=
\begin{bmatrix}  1 &  0   \\ 0   &  -1 \end{bmatrix}
\, ,
\end{align}
and used the well known property $\sigma_i \sigma_j=\delta_{ij} + i \epsilon_{ijk}\sigma_k$. Moreover, $\mathcal{\hat{M}}^2=0$

Boundary conditions are only mildly affected by these steps. After canonical normalization, the parameter $\lambda$ drops out of $\tilde{S}$, as $\phi'$ and $\psi'$ acquire identical dimensions. The subsequent analytic continuation renders $\tilde{S}$ symmetric rather than of the form $\tilde{S}_{\lambda=1}$ that we would obtain from Eq.~\eqref{eq:equivalent_2_order_problem_matrix_form_Kappa}. Thus, the number of free parameters in these matrices is the same and the BCs can be mapped into each other. Since the heat kernel will be computed for the operator \eqref{eq:D_rotated_continued}, we discuss the boundary conditions in more detail only for this case.

To set 
$
\omega_{\mathcal{\hat{D}}}
=
\langle \Psi'_1, \Phi'_2 \rangle
-
\langle \Psi'_2, \Phi'_1 \rangle
=
0
\, ,
$
we choose boundary conditions of the form
\begin{align}
\Phi'=
S
\Psi'
\, ,
\qquad S=S^T
=
 \begin{bmatrix} s_1  & s  \\ s& s_{2} \end{bmatrix}
 \, ,
\end{align}
which in components read
\begin{align}\label{eq:BC_general_form_rotaed_second_order_formulation}
(\partial_n\phi'=s_{1}\phi' + s\psi' \, , \quad \partial_n\psi'=s\phi' + s_{2}\psi')|_{\partial_M}
\, .
\end{align}
Notice that before analytic continuation we would simply have  $-s\phi'$ in place of $s\phi'$ in the second equation.
A particularly simple choice is to consider one of the nine combinatons of Dirichlet (D), Neumann (N), and Robin (R) boundary conditions valid for a single $\Box$ operaror. These correspond to the choices
\begin{align}\label{eq:choices_S_combinations_DNR}
(DD): \, &s_1=s_{2}=0, \, s \to \infty \, , & (DN):  \, &s=s_2=0, \, s_{1} \to \infty\, , & (ND): \, &s=s_{1}=0, \, s_{2} \to \infty\, ,  \\
(NN): \, &s=s_{1}=s_{2}=0 \, , & (DR): \, &s=0, \, s_{1} \to \infty   \, , & (RD): \, &s=0, \,  s_{2} \to \infty  \, , \\
(NR): \, &s=s_1=0\, , & (RN): \, &s=s_2=0\, , & (RR): \, &s=0 \, ,
\end{align}
where the notation $(B1,B2)$ means that we impose $B1$ on $\phi$ and $B2$ on $\psi$, respectivelty. For example, $(ND): (\partial_n\phi=0, \psi=0)$. However, also BCs mixing $\phi'$ and $\psi'$ and their derivatives are allowed.

For later convenience let us introduce the projectors
\begin{align}
\Pi_{-}
=
\begin{bmatrix} 0  &0   \\0  & 1  \end{bmatrix}
\, ,
\quad
\Pi_{+}
=
\begin{bmatrix} 1  &0   \\0  & 0  \end{bmatrix} 
\, ,
\end{align}
in terms of which we can compactly write
\begin{align}\label{eq:equivalent_2_order_problem_matrix_form_K}
(
\mathcal{B}_i
\Pi_{-}
\Psi'|_{\partial M}
=
0
\, ,
\quad
\mathcal{B}_j
\Pi_{+}
\Psi'|_{\partial M}
=
0
)
\, ,
\,
\forall \, i,j=1,2,3,4
\end{align}
where the boundary operators $\mathcal{B}_{i=1,2,3,4}$ can be one of the following
\begin{align}
\mathcal{B}_1
=
\begin{bmatrix} 1 & 0  \\ 0 & 1 \end{bmatrix}
\, ,
\quad
\mathcal{B}_2
=
\begin{bmatrix} \partial_n & 0  \\ 0 & \partial_n \end{bmatrix}
\, ,
\quad
\mathcal{B}_3
=
\begin{bmatrix} \partial_n -S & 0  \\ 0 & \partial_n -S\end{bmatrix}
\, ,
\quad
\mathcal{B}_4
=
\begin{bmatrix} \partial_n -S' & 0  \\ 0 & \partial_n -S' \end{bmatrix}
\, .
\quad
\end{align}

For clarity, let us now find explicitly the mapping between the boundary conditions in the four order formulation and the rotated (and analytic continued) thoery. We have
\begin{align}
\mathcal{B}_i
\Pi_{\pm}
\Psi'
=
\mathcal{B}_i
\Pi_{\pm}
R^{-1}\Psi
=
R^{-1}\left[
\mathcal{B}_i
\left(
R
\Pi_{\pm}
R^{-1}\Psi
\right)
\right]
=
0
\quad
\Rightarrow
\quad
\mathcal{B}_i
\left(
R
\Pi_{\pm}
R^{-1}\Psi
\right)
=
0
\, .
\end{align}
For $R=R^{-1}(\lambda,\pi/4)$ we get
\begin{align}
R
\Pi_{\pm}
R^{-1}\Psi
=
\frac{1}{2}
\begin{bmatrix} \phi \mp \lambda\psi   \\      \psi  \mp \frac{1}{\lambda}\phi  \end{bmatrix}
\, .
\end{align}
Thus, setting $i=j$ in Eq.~\eqref{eq:equivalent_2_order_problem_matrix_form_K} imposes the same boundary condition on both $\phi$ and $\psi$. In particular, the BCs in Eq.s~\eqref{eq:good_BC_flat_space_auxiliary_field} correspond to
\begin{align}
(
\mathcal{B}_i
\Pi_{-}
\Psi'|_{\partial M}
=
0
\, ,
\quad
\mathcal{B}_i
\Pi_{+}
\Psi'|_{\partial M}
=
0
)
\, ,
\,
 \,i=1,2
\end{align}
respectively. Ultimately, these coincide with the BCs \eqref{eq:BCs_flat_space_biharmonic2} of the original fourth-order formulation according to
\begin{align}\label{eq:map_no_parameters_BCs}
(\mathcal{B}_1,\mathcal{B}_1) \,\to\, (\Phi^0=0, \Phi^2=0)  \, , \qquad  (\mathcal{B}_2,\mathcal{B}_2) \,\to\, (\Phi^1=0, \Phi^3=0)   \, ,
\end{align}
reported here for readibility. These BCs were those independent of massive parameters and which remain admissible in the second-order formulation.

On the other hand, taking $i \neq j$ imposes mixed conditions on $\Psi$. For example, the chiral BC of Eq.~\eqref{eq:off_diagonal_BC_flat_space_auxiliary_field1} correspond to
\begin{align}
(
\mathcal{B}_1
\Pi_{-}
\Psi'|_{\partial M}
=
0
\, ,
\quad
\mathcal{B}_3
\Pi_{+}
\Psi'|_{\partial M}
=
0
)
\, ,
\end{align}
or, equivalently, with $\mathcal{B}_1$ with $\mathcal{B}_3$ exchanged. In the original fourth-order formulation, these correspond to the conditions 
\begin{align}\label{eq:mixed_BC_original_formulation}
(\Phi^0=\alpha \Phi^2, \Phi^1=-\alpha(\Phi^3 + \beta \Phi^0))
\, ,
\end{align}
we found in Eq.~\eqref{eq:chiral_BC_orginal_formulation}. Exchanging  $\mathcal{B}_1$ with $\mathcal{B}_3$ has just the effect of switching the sing in Eq.~\eqref{eq:chiral_BC_orginal_formulation}.
In a similar manner taking $\mathcal{B}_1$ and $\mathcal{B}_2$, yields the BC of Eq.s~\eqref{eq:off_diagonal_BC_flat_space_auxiliary_field1} for $s=0$ and correspond to \eqref{eq:mixed_BC_original_formulation} with $\beta=0$.
Instead, the choice
\begin{align}
(
\mathcal{B}_3
\Pi_{-}
\Psi'|_{\partial M}
=
0
\, ,
\quad
\mathcal{B}_4
\Pi_{+}
\Psi'|_{\partial M}
=
0
)
\, ,
\end{align}
yields Eq.~\eqref{eq:off_diagonal_BC_flat_space_auxiliary_field3} and corresponds to the linear combinations of Eq.~\eqref{eq:general_BC_biharmonic_llinear_combinations}. The last possibility, which is to consider $\mathcal{B}_2$ and $\mathcal{B}_3$, corresponds to Eq.~\eqref{eq:off_diagonal_BC_flat_space_auxiliary_field3} with $S=0$.

\subsection{Heat kernel for $\Box^2$ with Boundaries}\label{subsect:heat_kernel_flat_space}

The heat kernel $K(s;x,y)$ for the second derivative formulation \eqref{eq:D_rotated_continued} can be easily computed exactly. 
On a manifold without boundaries, $\hat{K}(s;x,y)$ satisfies the heat equation 
\begin{align}\label{eq:heat_equations}
(\partial_s + \mathcal{\hat{D}})\hat{K}(s;x,y) 
\, ,
\quad
\hat{K}(0;x,y)=\hat{1}\delta(x,y)
\, ,
\end{align}
where $\mathcal{\hat{D}}$ is the Heassian operator derived from some action $S[\varphi]$, see \cite{Schwinger:1951nm,DeWitt:1964mxt} or \cite{Vassilevich:2003xt} for an extensive review. In general, $\hat{K}(s;x,y)={K}^A{}_B(s;x,y)$ is a matrix acting on the space where the multicomponent field $\varphi=\varphi^A$ lives. 

\mycomment{
In the case of non-empty boundary we have to impose appropriate boundary conditions on the heat kernel.
Consider the eigenvalue problem for the Hessian operator
\begin{align}\label{eq:eigenvalue_problem}
\hat{D} u_\lambda(x)
=
\lambda  u_\lambda(x)
\, ,
\quad
u_\lambda(x)
=
\begin{bmatrix} 
u^1_\lambda(x)   \\[0.1em] 
u^2_\lambda(x) \end{bmatrix}
\, ,
\end{align}
where of course $\lambda$ and $u_\lambda(x)$ are eigenvalues and eigenfunctions respectively.
Then, the heat kernel admits the spectral decomposition 
\begin{align}\label{eq:spectral_decomposition}
\hat{K}(s;x,x')=\int_\lambda {\rm e}^{-s\lambda}u_\lambda(x)u^\dagger_\lambda(x')
\sim
\int_\lambda
\begin{bmatrix} 
u^1_\lambda(x) u^{1*}_\lambda(x') & u^1_\lambda(x) u^{2*}_\lambda(x')  \\[0.2em] 
u^2_\lambda(x) u^{1*}_\lambda(x') & u^2_\lambda(x) u^{2*}_\lambda(x') 
\end{bmatrix}
\, .
\end{align}
Thus, imposing a boundary condition on $u^1_\lambda(x)$ translates into the same condition on the first row (or column because of self-adjointness) of $ K(s;x,x')$, whereas a condition on $u^2_\lambda(x)$ imposes it on the second row.
}

For our operator, the simplest way to compute $\hat{K}$ is to consider a factorized solution
\begin{align}\label{eq:factorized_ansatz}
\hat{K}(s;x,x')={\rm e}^{-s\mathcal{\hat{M}}} \hat{K}_0(s;x,x')
\, .
\end{align}
Then, for $\mathcal{\hat{D}}=\mathcal{\hat{D}}_0 + \mathcal{\hat{M}}$, it is simple to see that Eq.~\eqref{eq:heat_equations} is satisfied for $K_0$ such that
\begin{align}
\partial_s \hat{K}_0+ \hat{D}_0\hat{K}_0 =0 \, ,
\quad
\hat{K}_0(0;x,x')=\hat{1}\delta(x,x') 
\mycomment{& \partial_s \hat{f} +  \mathcal{\hat{M}}  \hat{f} =0\, ,
\quad
\hat{f}(0)=\hat{1}\\}
\, .
\end{align}
The solution has the diagonal form
\begin{align}\label{eq:spectral_decomposition}
\hat{K}_0(s;x,x')
=
\begin{bmatrix} 
K^{B1}_\Box(s;x,x')   & 0 \\[0.2em] 
0                             & K^{B2}_\Box(s;x,x')
\end{bmatrix}
\, .
\end{align}
where $K_{\Box}^{Bi}$ is the single field heat kernel with any of the possible boundary conditions, i.e., Dirichlet, Neumann or Robin boundary conditions.

Thanks to the properties of $\mathcal{\hat{M}}$, it is simple to compute
\begin{align}
 {\rm e}^{-s\mathcal{\hat{M}}} 
 =
1
+
\frac{sm^2}{2}(\sigma_3 - i \sigma_1)
=
\begin{bmatrix}  
1 + \frac{sm^2}{2}  &   - i \frac{sm^2}{2}   \\ 
-i \frac{sm^2}{2}     &   1-\frac{sm^2}{2} 
\end{bmatrix}
\, .
\end{align}
Thus, for the trace we obtain
\begin{align}\label{eq:trace_HK_boxSquared_result}
{\rm Tr} \, \hat{K}^{B1,B2}_{\Box^2}(s)
=
\int d^dx   \left[
K^{B1}_\Box(s;x,x)  \left( 1 + \frac{sm^2}{2} \right)
+
K^{B2}_\Box(s;x,x)  \left( 1 - \frac{sm^2}{2} \right)
\right]
\, ,
\end{align}
where the $i$-dependence ddisappears. This equation yield the trace of the heat kernel of $\Box^2$ under a wide range of non-standard boundary conditions according to the mapping established at the end of the previous section.
From Eq.~\eqref{eq:trace_HK_boxSquared_result} we obtain that in the coincidence limit the heat kernel coefficients take the form
\begin{align}\label{eq:coincindence_limit_HK_boxSquared_result}
a_n(f,\Box^2, (B1,B2))
=
a_n(f,\Box, B1)
+
a_n(f,\Box, B2)
+
\frac{m^2}{2}
\left(
a_{n-2}(f,\Box, B1)
-
a_{n-2}(f,\Box, B2)
\right)
\, .
\end{align}
Interestingly, under the conformal boundary conditions we will derive in Sect.~\ref{sect:anomaly_curved_space} (see Eq.~\eqref{eq:conf_BC_in_flat_space}), the dependence on the mass parameter $m$ drops out entirely.
We also emphasize that Eqs.~\eqref{eq:coincindence_limit_HK_boxSquared_result} and \eqref{eq:trace_HK_boxSquared_result} for the $\Box^2$ operator remain valid on curved backgrounds. Furthermore, the right hand-side of \eqref{eq:coincindence_limit_HK_boxSquared_result} is in agreement with the known heat kernel coefficients for operators of the form $\mathcal{\hat{D}}=-{\hat{1} }  \Box +   \hat{E}$ with mixed boundary conditions which can be found in \cite{Vassilevich:2003xt}.

Using these results we now compute the effective action. As an example, we perform this computation for $(D,N)$ boundary conditions.
The one-loop effective action $\Gamma$ is related to the heat trace through
\begin{align}
\Gamma^{B1,B2}_{eff,\Box^2}
=
-\frac{1}{2}
\int_{\epsilon^2}^{L^2}  
\frac{ds}{s}
{\rm Tr} \, {K}^{B1,B2}_{\Box^2}(s)
\, ,
\end{align}
where $\epsilon$ and $L$ are an UV and an infrared cur off, respectively. In flat space it is simple to evaluate $\Gamma$ explicitly thank to  Eq.~\eqref{eq:trace_HK_boxSquared_result}. 
We consider the manifold $M=\mathbb{R}^{d-1} \times \mathbb{R}^+$ with boundary $\partial M=\{z=0\}$.
For Neumann (Dirichlet) BCs at $z=0$
\begin{align}
\left. \partial_z K^{N}_\Box(s;x,x') \right|_{z=0}
=
0
\, ,
\quad
\left. K^{D}_\Box(s;x,x') \right|_{z=0}
=
0
\, ,
\end{align}
the heat kernel can be find by using the image method to be
\begin{align}
K^{N(D)}_\Box(s;x,x')
=
\frac{1}{(4\pi s)^{d/2}}
\left(
e^{-\frac{|x-x'|^2}{4s}}
\pm
e^{-\frac{|x-x'_{im}|^2}{4s}}
\right)
\, 
,
\end{align}
where the image $x'_{im}=(\vec{x'}_\parallel, -z')$. A simple integration of the coincidence limit of the previous equation shows that
\begin{align}\label{eq:effA_ND}
\Gamma^{N(D)}_{eff}
=
-
\frac{V_{d-1}}{2}
\int_{\epsilon^2}^{\infty}  
\frac{ds}{s}
\int_{\mathbb{R}^+} dz 
\frac{1}{(4\pi s)^{d/2}}
\left(
1
\pm
e^{-\frac{z^2}{s}}
\right)
=
-
\frac{V_M}{(4\pi)^{d/2}}\frac{1}{d \epsilon^d}
\mp
\frac{V_{d-1}}{4(4\pi)^{\frac{d-1}{2}}}\frac{1}{(d-1) \epsilon^{d-1}}
\, ,
\end{align}
where $V_{d-1}=\int_{\mathbb{R}^{d-1}}d\vec{x}_\parallel$, $V_{M}=\int_{\mathbb{R}^{d-1}}\int_{\mathbb{R}^+}d\vec{x}_\parallel dz$ and we send $L \to \infty$.
Putting together Eq.~\eqref{eq:effA_ND} and Eq.~\eqref{eq:trace_HK_boxSquared_result}, we get
\begin{align}
\Gamma^{D,N}_{eff}
=
-
\frac{2V_M}{(4\pi)^{d/2}}\frac{1}{d \epsilon^d}
+
\frac{ V_{d-1} }{2}
\int_{\epsilon^2}^{\infty}  
\frac{ds}{s}
\int_{\mathbb{R}^+}
\frac{sm^2}{(4\pi s)^{d/2}}
e^{-\frac{z^2}{s}}
\, .
\end{align}
For $d > 3$ the last integral yields
\begin{align}
\int_{\epsilon^2}^{\infty} 
ds
\int_{0}^{\infty} dz 
\frac{     e^{-\frac{z^2}{s}}   }{s^{d/2}}
=
\frac{ \pi^{1/2}  }{ d-3 }  
\frac{1}{\epsilon^{d-3}}
\, ,
\end{align}
and in $d=4$ the full result is
\begin{align}
\Gamma^{D,N}_{eff}
=
-
\frac{V_M}{2(4\pi)^{2}}\frac{1}{\epsilon^4}
+
\frac{ m^2 V_3 }{32\pi^{3/2}}\frac{1}{\epsilon}
\, .
\end{align}
The first term is the usual bulk divergence which renormalize the cosmological constant $\Lambda$, while the second is a boundary divergence renormalizing a boundary tension $\tau$
\begin{align}
\Lambda_B = \Lambda_R + \frac{1}{2(4\pi)^2}\,\frac{1}{\epsilon^4},
\qquad
\tau_B = \tau_R - \frac{m^2}{32\pi^{3/2}}\,\frac{1}{\epsilon}
\, ,
\end{align}
where the suffix $B$ and $R$ stand for bare and renormalized, respectively.

A simple fact to keep in mind is that, since our field-reparametrization does not involve any differential operator, the Jacobian in the path integral is trivial and does not contribute to the divergences. In fact, we have
\begin{align}
1=\int D\Psi e^{-\frac{1}{2}\Psi^T\Psi} 
=
\int D\Psi' J e^{-\frac{1}{2}\Psi'^T(R^TR)\Psi'} 
\Rightarrow
J \propto \det(R^TR)^\frac{1}{2} \, , \qquad 
\text{where}
\qquad
R
=
\frac{1}{\sqrt{2}}\begin{bmatrix}  1 & \lambda   \\ -\frac{1}{\lambda} & 1   \end{bmatrix}
\, .
\end{align}

\section{Auxiliary Fields in Higher-Derivative Boundary CFTs: Curved Space}\label{sect:anomaly_curved_space}

\subsection{Auxiliary Field Formalism and Conformal Invariance}\label{subsect:weyl_invariance_auxiliary_field}

We consider a manifold $M$ with a co-dimension one boundary $\partial M$. In curved space, the Weyl invariant extension of the original fourth-order formulation is
\begin{align}\label{eq:paneitz_formulation}
\tfrac{1}{2}
\int_M d^4  x \sqrt{ g} \, 
\phi
\Delta_4
\phi
\, , 
\qquad
\qquad
\Delta_4= 
\Box^2
+
\nabla_\mu
  \left(V^{\mu\nu}
\nabla_\nu
\right)
\equiv
\Box^2
+
\nabla_\mu
\left(
\left(
2 R^{\mu\nu}
-
\frac{2}{3}g^{\mu\nu}R
\right)
\nabla_\nu
\right)
\, ,
\end{align}
where $\Delta_4$ is the Fradkin-Tseytlin-Paneitz operator \cite{Fradkin:1982xc,Riegert:1984kt,paneitz} and $V^{\mu\nu}$ is defined implicitly.
Introducing an auxiliary-field the action 
\begin{align}\label{eq:weyl_invariant_second_order_formulation}
 S(\phi,\psi)
 =
 \tfrac{1}{2}
 \int_M d^4 x  \sqrt{g}
\Psi^T \hat{{D}}_2 \, \Psi
=
\tfrac{1}{2}
\int_M d^4  x \sqrt{ g} 
\left(
  \phi \Box \psi
  + \psi \Box \phi
  - \psi^2
    +
  \phi
  \nabla_\mu
  \left(V^{\mu\nu}
\partial_\nu
\phi
\right)
\right)
\, 
\end{align}
yields \eqref{eq:paneitz_formulation} upon integrating out $\psi$. We have defined the $2\times 2$ operator 
\begin{align}\label{eq:D2_operator_curved_space}
\hat{D}_2 \;=\;
\begin{bmatrix}
  \begin{aligned}
      \nabla_\mu      \left(      V^{\mu\nu}   \partial_\nu      \right)
  \end{aligned}
  &
  \begin{aligned}
    \Box 
  \end{aligned}
  \\
  \begin{aligned}
    \Box 
  \end{aligned}
  &
  -1
\end{bmatrix}
\, 
\end{align}
acting on $\small{\Psi=\begin{bmatrix} \phi \\ \psi \end{bmatrix}}$. Notice that, naively, one has $\det \hat{D}_2 \sim \det \Delta_4$.

The equivalence of the two formulations can be established by using the equations of motion. From Eq.~\eqref{eq:weyl_invariant_second_order_formulation}, under suitable BCs we are about to discuss, we obtain
\begin{align}\label{eq:eom_weyl_invariant_auxiliary_fields}
\psi=\Box\phi \, , \qquad   \Box\psi +    \nabla_\mu      \left(      V^{\mu\nu}   \partial_\nu  \phi    \right) = 0 \, .
\end{align}
where the first equation defines the auxiliary field while the second reproduces the equation of motion derived from \eqref{eq:paneitz_formulation}.

For a term such as $\nabla_\mu ({V}^{\mu\nu}\nabla_\nu)$ we have
\begin{align}
 \int_M d^4 x  \sqrt{g} 
 \phi_1 \nabla_\mu ({V}^{\mu\nu}\nabla_\nu \phi_2) 
=
 \int_M d^4 x  \sqrt{g}
\phi_2  \nabla_\mu ({V}^{\mu\nu}\nabla_\nu \phi_1) 
+
\int_{\partial M} d^3 y \sqrt{h} 
\left( 
\phi_1 {V}^{n\mu} \partial_\mu\phi_2  -  \phi_2 {V}^{n\mu}    \partial_\mu \phi_1 
\right) \, ,
\end{align}
where ${V}^{n\nu}=n_\mu{V}^{\mu\nu}$, $n$ is the outward pointing normal vector, $h$ is the induced metric on the boundary, and $y^a$ are the intrinsic coordinates on the boundary. Up to the global $\tfrac{1}{2}$ factor, the boundary terms descending from \eqref{eq:weyl_invariant_second_order_formulation} can be written as 
\begin{align}\label{eq:boundary_terms_weyl_invariant_bulk_auxiliary_fields}
\omega_{\hat{{D}}_2 }
=
\int_{\partial M} d^3 y  \sqrt{h} \;
&\Big(
\phi_1 ( \partial_n \psi_2  +{V}^{n\nu} \partial_\nu \phi_2   )
- 
 \phi_2 ( \partial_n \psi_1 + {V}^{n\nu}    \partial_\nu \phi_1   ) 
 +
\psi_1\partial_n\phi_2 \, + \\ \nonumber
&
-
\psi_2 \partial_n \phi_1
\Big)
=
\left\langle 
W_1, JV_2 
\right\rangle
-
\left\langle
W_2, J V_1 
\right\rangle
\, ,
\end{align}
where we defined 
\begin{align}
W 
=
\begin{bmatrix}\phi    \\ \psi   \end{bmatrix}
\, ,
\qquad
V
=
\begin{bmatrix}  \partial_n \phi  \\  \partial_n \psi  +  {V}^{n\nu} \partial_\nu \phi  \end{bmatrix}
\, ,
\quad
J
=
\begin{bmatrix} 0 & 1  \\ 1 & 0 \end{bmatrix}
\, .
\end{align}
In passing through, we notice that since
\begin{align}
 \int_M d^4 x  \sqrt{g}
\phi_1 \Box^2   \phi_2
=
 \int_M d^4 x  \sqrt{g}
\phi_2 \Box^2\phi_1
+
\int_{\partial M} d^3 y \sqrt{h} 
\left( 
\phi_1 \partial_n \Box \phi_2  
-\partial_n\phi_1   \Box   \phi_2     
-  \phi_2 \partial_n \Box \phi_1  
+\partial_n\phi_2   \Box   \phi_1    
 \right) \, ,
\end{align}
the boundary terms of the fourth and second order formulations are mapped into each other once we go on-shell with the auxiliary field.

As in flat space, the matrix defining the boundary conditions setting $\omega_{\hat{{D}}_2 }$ to zero should be $J$-symmetric. Thus, we consider boundary conditions such as %eq:J_symmetric_S
\begin{align}
V
=
S\Psi
\, , \qquad
S
=
\begin{bmatrix} s & s_{12}  \\ s_{21} & s \end{bmatrix}
\, ,
\end{align}
or equivalently
\begin{align}\label{eq:BC_curved_space_second_order}
\left. \Big( \partial_n \phi = s\phi + s_{12}\psi  \, , \quad  \partial_n \psi + {V}^{n\nu} \partial_\nu \phi  =s_{21}\phi + s\psi   \Big) \right|_{\partial M}
\, .
\end{align}
By choosing $s_{12}=s_{21}=0$, and $s \to \infty$, we haves the Dirichlet conditions $( \phi = 0,   \psi=0   )|_{\partial M}$.
It is simple so see they set $\omega_{\hat{{D}}_2 }=0$. On the other hand, in cruved space it is not possible to choose simultaneously Neumann on $\phi$ and $\psi$, as $\partial_n \psi$ and $ {V}^{n\nu} \partial_\nu \phi$ necessarily come with the same coefficient. Another choice that would be algebrically allowed is $(\phi = 0,   \partial_n\phi=0   )|_{\partial M}$. However, as we previously discussed, this would mean to impose two conditions on a second order system.

%\subsection{Conformal and Weyl invariance }

Let us now consider Weyl invariance. 
We take the metric to transform in the standard way $g'_{\mu\nu}=e^{2\sigma}g_{\mu\nu}$. Thus, the infinitesimal Weyl transformation of the Christoffel connection is
\begin{align}
\delta^W_\sigma \Gamma^\rho{}_{\nu\mu}
=
\delta^\rho_\nu \partial_\mu\sigma
+
\delta^\rho_\mu \partial_\nu\sigma
-
g_{\nu\mu}\partial^\rho\sigma \, .
\end{align}
Our convention for covariant differentiation is that the last index is the directional one. The Weyl transformation of the normal vector is
\begin{align}
n^\nu \to n'^\nu=e^{-\sigma}n^\nu\, , 
\end{align}
as can be seen by requiring $1=g'_{\mu\nu}n'^\mu n'^\nu=e^{2\sigma}g_{\mu\nu}n'^\mu n'^\nu$. The induced metric $h_{ab}=e_a{}^\mu e_b{}^\nu g_{\mu\nu}$ transforms as $h'_{ab}=e^{2\sigma}|_{\partial M} h_{ab}$. It is simple to see that
\begin{align}\label{eq:weyl_transformations_extrinsic_curvature}
\mycomment{& K_{\mu\nu} = h_{\mu}{}^{\alpha}   h_{\nu}{}^{\beta}   \nabla_\alpha n_\beta
\to
e^{\sigma} h_{\mu}{}^{\alpha}   h_{\nu}{}^{\beta} \left[ \nabla_\alpha n_\beta
-
n_\alpha \partial_\beta \sigma
+ g_{\alpha\beta} \partial_n \sigma
\right] 
=
e^{\sigma} \left[ K_{\mu\nu} 
+ h_{\mu\nu} \partial_n \sigma
\right] \, , \\
&} K_{ab} 
\to
e^{\sigma} \left[ K_{ab} 
+ h_{ab} \partial_n \sigma
\right] \, , \qquad 
 K
\to
e^{-\sigma} \left[ K
+ (d-1)\partial_n \sigma
\right]
\, , \qquad 
 \hat{K}_{ab} 
\to
e^{\sigma} \hat{K}_{ab} 
\end{align}
where $\hat{K}_{ab}={K}_{ab}-\frac{h_{ab}}{d-1}K$ is the traceless part of the extrinsic curvature ${K}_{ab}=e_a{}^\mu e_b{}^\nu \nabla_\mu n_\nu$.
%It is convenient to recall the decomposition $\nabla_\mu n_\nu = K_{\mu\nu} + n_\mu a_\nu$, where $a_\nu=n^\mu \nabla_\mu n_\nu$ is the purely spatial acceleration vector, and $K_{\mu\nu} = h^\alpha_\mu h^\beta_\nu \nabla_\alpha n_\beta$. 
\begingroup
\setlength{\abovedisplayskip}{6pt}
\setlength{\belowdisplayskip}{6pt}
\setlength{\abovedisplayshortskip}{0pt}
\setlength{\belowdisplayshortskip}{3pt}

 A field with Weyl weight $w_{\phi^i}$ transforms homogeneusly as
\begin{align}\label{eq:Weyl_transformation_primary_fields}
\phi^i \to \phi'^i= e^{w_{\phi^i}\sigma}\phi^i \, .
\end{align}
For a four derivative scalar $\phi$ the weight is zero $w_{\phi}=0$.
Since the equations of motion dictates $\psi = \Box \phi$, the field $\psi$ is not a conformal primary and we take its Weyl variation to be the same of $\Box \phi $. We choose
\begin{align}\label{eq:Weyl_transformation_psi}
\psi \;\to\; \psi'= e^{-2\sigma}
\left(
  \psi
  + (d-2)\,\partial_\mu \sigma\, \partial^\mu \phi
\right) 
\, ,
\end{align}
which preserves the defining relation $\psi=\Box\phi$ under Weyl transformations. 
With this choices---for $\partial M = \emptyset$---the second-order action \eqref{eq:weyl_invariant_second_order_formulation} is already Weyl invariant. The easiest way to see this is to notice that
\begin{align}\label{eq:off_shell_weyl_invariance_second_order}
 S(\phi,\psi)
& =
\tfrac{1}{2}
\int_M d^4  x \sqrt{ g} 
\left(
  \phi \Box \psi
  + \psi \Box \phi
  - \psi^2
    +
  \phi
  \nabla_\mu
  \left(V^{\mu\nu}
\partial_\nu
\phi
\right)
\right)\\ \nonumber
 &=
 \tfrac{1}{2}
 \int_M d^4 x  \sqrt{g}
\left(
\phi \Delta_4 \phi
-
X^2
\right)
\, ,
 \end{align}
where
\begin{align}
X=\psi -\Box\phi \, , \qquad  X \to X'=e^{-2\sigma}X
\, .
 \end{align}

\endgroup

In the presence of a non empty boundary, Eq.~\eqref{eq:off_shell_weyl_invariance_second_order} holds only up to the boundary term
\begin{align}
 B(\phi,\psi)
 =
 - \tfrac{1}{2}
 \int_{\partial M} d^3 y  \sqrt{h}
\left(
X\partial_n \phi
-
\phi \partial_n X
\right)
\, ,
 \end{align}
which however breaks conformal invariance. Invariance may be restored \emph{off-shell} by adding 
$ B_1(\phi,\psi)
\propto
\int_{\partial M} d^3 y  \sqrt{h} \,
\phi K X$, where $K$ is the trace of the extrinsic curvature. The role of this term is precisely to cancel the boundary contributions coming from $\delta_\sigma S(\phi,\psi)$. In this way, the action
\begin{align}\label{eq:weyl_invariant_auxiliary_fields_boundary}
S(\phi,\psi)
 &=
\tfrac{1}{2}
\int_M d^4  x \sqrt{ g} 
\left(
  \phi \Box \psi
  + \psi \Box \phi
  - \psi^2
    +
  \phi
  \nabla_\mu
  \left(V^{\mu\nu}
\partial_\nu
\phi
\right)
\right)
-
\tfrac{1}{3}
\int_{\partial M} d^3 y  \sqrt{h}
\phi K X \\ \nonumber
&= 
\left(\Psi, \hat{{D}}_2 \, \Psi \right)
+
\left\langle \Psi, B \Psi \right\rangle
\, ,
\end{align}
is Weyl invariant.  We recall that we defined $(, \, )$ and $\langle, \, \rangle $ to be scalar product over $M$ and $\partial M$, respectively. The operator $\hat{{D}}_2$ is the same as in Eq.~\eqref{eq:D2_operator_curved_space} and we introduced
\begin{align}
B
=
\begin{bmatrix} -K\Box\,\,\, & \tfrac{K}{2}  \\ \tfrac{K}{2} & 0 \end{bmatrix}
\, .
\end{align}

That the action \eqref{eq:weyl_invariant_auxiliary_fields_boundary} is Weyl invariant is clear when we write it as
\begin{align}\label{eq:weyl_invariant_auxiliary_fields_boundary_X}
 S(\phi,\psi)
=
\tfrac{1}{2}
 \int_M d^4 x  \sqrt{g}
\left(
\phi \Delta_4 \phi
-
X^2
\right)
+
\tfrac{1}{2}
 \int_{\partial M} d^3 y  \sqrt{y}
\left(
X\partial_n \phi
-
\phi (\partial_n X + \tfrac{2}{3} KX)
\right)
\, ,
\end{align}
which is manifestly invariant---also for any $d \geq 4$ with minor modifications. Notice that is action is now \emph{linear} in $X$.

\subsubsection{Conformal Boundary Conditions and Auxiliary Fields}\label{subsect:conf_bc_auxiliary_fields}

Conformal boundary operators for $\Delta_4$ are known in the mathematical literature and can be found in \cite{Case:Paneitz_Bop}. In that work, it has been shown that 
\begin{align}\label{eq:paneitz_boundary_terms}
\omega_{\Delta_4}
=
\int_{\partial M} d^{3} x  \sqrt{h} 
\left(
\Phi_{2,c}^0 \Phi_{1,c}^3 
+
\Phi_{2,c}^1 \Phi_{1,c}^2
-
\Phi_{1,c}^0 \Phi_{2,c}^3 
-
\Phi_{1,c}^1 \Phi_{2,c}^2
\right)
\, ,
\end{align}
where
\begin{subequations}
\begin{align}
\Phi^0_c= \, & \phi|_{\partial M} \,,  \label{subeq:0th_order_paneitz_BC}\\
\Phi^1_c= \, &\partial_n \phi|_{\partial M} \,, \label{subeq:1th_order_paneitz_BC}\\[4pt]
\Phi^2_c= \, &\left( \nabla_n \partial_n +  \tfrac{1}{3}K \partial_n - D^2 \right)\phi\Big|_{\partial M} \, , \label{subeq:2th_order_paneitz_BC}\\[4pt]
\Phi^3_c= \, &\left(-\partial_n \Box - 2D^2\partial_n 
+ 2 \hat{K}^{ab} D_a \partial_b 
- \tfrac{2}{3}K D^2
+ \tfrac{2}{3}\partial_a K\partial^a
+ \mathcal{R}\partial_n \right)\phi\Big|_{\partial M} \, ,  \label{subeq:3th_order_paneitz_BC}
\end{align}
\end{subequations}
with
\begin{equation}
\mathcal{R} 
=
\tfrac{1}{2}R^{(3)} - 2P_{\mu\nu}n^\mu n^\nu 
- \tfrac{1}{3}K^2 + \tfrac{1}{2}\hat{K}_{ab}\hat{K}^{ab}.
\end{equation}
are conformally invariant. The tensor $P_{\mu\nu}=\frac{1}{2}\left( R_{\mu\nu} - \tfrac{1}{6}R g_{\mu\nu} \right)$ is the Schouten tensor. Using this results it is easy to obtain the conformal boundary conditions from the action in Eq.~\eqref{eq:weyl_invariant_auxiliary_fields_boundary_X}. In this way, in the auxiliary field formalism we have
\begin{align}\label{eq:auxiliary_boundary_terms}
\omega_{\hat{D}_2,B}
&=
\int_{\partial M} d^{3} x  \sqrt{h} \,
\Big(
\Phi_{2,c}^0  \left( \Phi_{1,c}^3  - \partial_n X_1 - \tfrac{2}{3} KX_1 \right)
+
\Phi_{2,c}^1 \left( \Phi_{1,c}^2 +X_1 \right) \, +  \Phi_{1,c}^0  \left( \Phi_{2,c}^3   - \partial_n X_2 - \tfrac{2}{3} KX_2  \right) \\ \nonumber
&
\qquad\qquad 
-
\Phi_{1,c}^1 \left( \Phi_{2,c}^2 + X_2\right)
\Big)\\
&\equiv
\int_{\partial M} d^{3} x  \sqrt{h} \,
\Big(
\Phi_{2,c}^0   \tilde{\Phi}_{1,c}^3
+
\Phi_{2,c}^1 \tilde{\Phi}_{1,c}^2 
-
\Phi_{1,c}^0   \tilde{\Phi}_{2,c}^3   
-
\Phi_{1,c}^1  \tilde{\Phi}_{2,c}^2 
\Big)
\, .
\end{align}
With the same strategy as before, to set $\omega_{\hat{D}_2,B}=0$ we can impose
\begin{align}\label{eq:general_BC_aux_fied}
({\Phi}_{c}^1=s{\Phi}_{c}^0 + s_{12}\tilde{\Phi}_{c}^2  \, , \quad \tilde{\Phi}_{c}^3=s_{21}{\Phi}_{c}^0 + s\tilde{\Phi}_{c}^2)
\, ,
\end{align}
and the conformal BC are 
\begin{align}\label{eq:BCs_curved_space_aux_fied}
(&{\Phi}_{c}^0=0, \Phi^1=0) \, , & (& \tilde{\Phi}_{c}^2=0, \tilde{\Phi}_{c}^3=0)\, , \\
 (&{\Phi}_{c}^0=0, \tilde{\Phi}_{c}^2=0)  \, , & (&\Phi^1=0, \tilde{\Phi}_{c}^3=0)   %\label{eq:BCs_flat_space_biharmonic2}
\, .
\end{align}

The conformal boundary conditions $\phi|_{\partial M}=0$  and $\partial_n\phi|_{\partial M}=0$ are unaltered in the auxiliary field formalism.\footnote{We notice that, since $\phi$ has weight zero, Robin conditions $(\partial_n\phi - S \phi)|_{\partial M}=0$ are incompatible with conformal symmetry. $S$ should transform as $S \to e^{-\sigma}S$, but no local scalar curvature polynomial transforming homogenously with weight $-1$ exists. On the other hand, away from $d=4$, $\delta^W_\sigma \partial_n \phi \propto (d-4)\phi \partial_n \sigma $ and is thus possible as usual to restore Weyl invariance by adding $K\phi$.}
However, as we discussed above, even if the pair
\begin{align}\label{eq:BC_curved_space_first_order_weyl_invariant}
\left(\phi|_{\partial M}=0, \quad\partial_n\phi|_{\partial M}=0\right)
\, ,
\end{align}
is conformal invariant, in the auxiliary formalism they are not consistent and thus we discard them. 

Second and third order boundary conditions are insted modifield by the presence of $\psi$ and $\partial_n \psi$, rispectively. By inserting the decomposition
 \begin{align}\label{eq:tg_normal_decomposition_box_phi}
\Box \phi
=
 D^2  \phi + K  \partial_n \phi+ \nabla_n \partial_n\phi 
\, ,
\end{align}
into $ \tilde{\Phi}_{c}^2$, the conformal boundary condition involving $\psi$ becomes
\begin{align}\label{eq:conf_BC_psi}
\left.\left(\psi - \tfrac{2}{3}K\partial_n\phi -  2  D^2\phi \right)\right|_{\partial M}=0
\, ,
\end{align}
up to the addition of the Weyl invariant term $\tilde{K}^2_{ij}$. 
This condition can also be constructed directly in an easy way. Let us consider the following one parameter family of Weyl covariant operators in dimension $d$
\begin{align}\label{eq:second_order_conf_cov_op_psi}
\psi - \tfrac{d-2-c}{d-1}K\partial_n\phi -  \tfrac{d-2+c}{d-3}  D^2\phi + c \nabla_n \partial_n \phi  =0
\, .
\end{align}
The ambiguity parametrized by $c$ can be eliminated if we insist to obtain the first of allowed scale invariant boundary condition \eqref{eq:good_BC_flat_space_auxiliary_field} in the flat space limit, which require setting $c=0$. This way, we obtain again \eqref{eq:conf_BC_psi} for $d=4$.
%Moreover, it is simple to verify that \eqref{eq:conf_BC_psi} matches Eq.~\eqref{subeq:2th_order_paneitz_BC} on-shell. 
Thus, we arrive at the boundary conditions 
\begin{align}\label{eq:BC_curved_space_second_order_weyl_invariant}
\left(\phi|_{\partial M}=0, \quad\left.\left(\psi - \tfrac{2}{3}K\partial_n\phi \right)\right|_{\partial M}=0\right)
\, ,
\end{align}
which are Weyl invariant, set $\omega_{\hat{D}_2,B}=0$, and in flat space yield
\begin{align}\label{eq:BC_curved_space_second_order_weyl_invariant_flat_space}
\left(\phi|_{\partial M}=0, \quad     \psi|_{\partial M}=0 \right)
\, ,
\end{align}
which is one of the conditions we used in the first part of the paper to compute the effective action.

Similarly, the boundary condition involving $\partial_n \psi$ is obtained from $ \tilde{\Phi}_{c}^3$ and can be written as
\begin{equation}
\left(-\partial_n \psi - 2D^2\partial_n 
+ 2 \tilde{K}^{ab} D_a \partial_b 
+ \tfrac{2}{3}K \nabla_n\partial_n\phi
+ \tfrac{2}{3}\partial_a K\partial^a\phi
+ \mathcal{R}'\partial_n\phi\right)\Big|_{\partial M} =0 \, ,
\end{equation}
where now
$
\mathcal{R}'
=
\tfrac{1}{2}R^{(3)} - 2P_{\mu\nu}n^\mu n^\nu 
+\tfrac{1}{3}K^2 + \tfrac{1}{2}\hat{K}_{ab}\hat{K}^{ab}.
$ Thus, we have the pair of conformal boundary conditions
\begin{align}\label{eq:BC_curved_space_third_order_weyl_invariant}
\left(\partial_n \phi|_{\partial M}=0, \quad\left.\left(\partial_n \psi - 2 \hat{K}^{ab} D_a \partial_b  - \tfrac{2}{3}K \nabla_n\partial_n\phi   - \tfrac{2}{3}\partial_a K\partial^a\phi\right)\right|_{\partial M}=0\right)
\, ,
\end{align}
which in the flat space limit becomes
\begin{align}\label{eq:BC_curved_space_third_order_weyl_invariant_flat_space}
\left(\partial_n \phi|_{\partial M}=0, \quad    \partial_n \psi|_{\partial M}=0 \right)
\, .
\end{align}

Finally we have the boundary conditions $(\tilde{\Phi}_{c}^2=0, \tilde{\Phi}_{c}^3=0)$, which in the flat space limit yields $(\psi|_{\partial M}=0, \partial_n\psi|_{\partial M}=0)$, which again are not admissible.

\subsubsection{Weyl Invariance and Field Space Rotation}

Naturally, Weyl invariance survives the field space rotations.
Consider the action $S(\phi_i; g)$ depending on the fields $\phi_i$ 
($i = 1, \dots, N$) and on the metric $g_{\mu\nu}$. 
Under a global rotation---or any linear transformation---acting as
\begin{align}
\phi_i = R_{ij} \phi'_j, \qquad 
\phi'_i = (R^{-1})_{ij} \phi_j,
\end{align}
the action becomes
\begin{align}
 S'(\phi'; g) = S(\phi(\phi'); g) = S(R\phi'; g).
\end{align}
Then, under a Weyl variation $\delta^W_\sigma$, we have
\begin{align}\label{eq:variation_linearly_transfromed_action}
\delta^W_\sigma S'(\phi'; g)
= \tfrac{\delta  S'}{\delta \phi'_i} \,\delta^W_\sigma \phi'_i
+ \tfrac{\delta S'}{\delta g_{\mu\nu}} \,\delta^W_\sigma g_{\mu\nu}
= \delta^W_\sigma S(\phi; g)
\, .
\end{align}
Thus, in particular, if the original action is Weyl invariant, so is the transformed action.
In the Eq.~\eqref{eq:variation_linearly_transfromed_action} we used that $R$ is a global transformation and the chain rule
\begin{align}\label{eq:weyl_variation_trasnformed_fields}
\delta_\sigma \phi'_i 
= R^{-1}_{ij} \, \delta_\sigma \phi_j
\, ,
\qquad
\frac{\delta S'}{\delta \phi'_i}
= \frac{\delta S}{\delta \phi_j} 
\frac{\delta \phi_j}{\delta \phi'_i}
= \frac{\delta S}{\delta \phi_j} R_{ji}
\, .
\end{align}
Moreover,  
%\begin{align}
${\small{
\tfrac{\delta \tilde S}{\delta g_{\mu\nu}}\delta_\sigma g_{\mu\nu}
= 
\tfrac{\delta S}{\delta g_{\mu\nu}}\delta_\sigma g_{\mu\nu} }}
$
%\end{align}
being the Weyl variation of the metric independent of the fields. Analogously, a Weyl invariant boundary condition will be rotated in a Weyl invariant boundary condition
\begin{align}
\delta_\sigma B'(\phi'_i; g)
=
\frac{\delta B(R_{ij}\phi_j; g)}{\delta \phi_k} \frac{\delta \phi_k}{\delta \phi'_l}\delta_\sigma \phi'_l
+
\frac{\delta B(R_{ij}\phi_j; g)}{\delta g_{\mu\nu} } \delta_\sigma g_{\mu\nu} 
=
\delta_\sigma B(\phi_i)
\, .
\end{align}
Thus, in our case, the flat space limit rotated action
\begin{align}
\int_M d^4  x
\left(
-\frac{1}{\lambda}\phi' \Box \phi'
-
 \frac{1}{2\lambda^2} \phi'^2
+
 \lambda \psi' \Box \psi'
-
\frac{1}{2} \psi'^2
+
\frac{1}{\lambda} \phi'\psi'
\right)
\, ,
\end{align}
where the parameter $\lambda$ is left arbitrary for generality, is conformally invariant if we take the rotated fields in curved space to transform as
\begin{align}
\delta^W_\sigma \phi' = -\lambda \delta^W_\sigma(  \Box \phi )
\, ,
\qquad
\delta^W_\sigma \psi' = \delta^W_\sigma(\Box \phi )
\, ,
\end{align}
as dictated by Eq.~\eqref{eq:weyl_variation_trasnformed_fields}.

\subsection{Boundary Charges in the Auxiliary Field Formalism}\label{subsect:boundary_charges_auxiliary}

In addition to the familiar $a$ and $b$ bulk charges, for four dimensional CFT the trace anomaly acquires two new boundary central charges $b_1$ and $c$. The integrated anomaly can be written as \cite{Fursaev:2015wpa,Herzog:2015ioa,Solodukhin:2015eca}
\begin{equation}\label{eq:trace_anomaly_with_boundaries}
16\pi^2\int_M d^4 x \sqrt{g} \langle T \rangle 
=
-a \chi(M)
+
 \int_M d^4 x \sqrt{g} b W^2
- 
 \int_M d^3 y \sqrt{h} \left( c\, \text{tr}\, \hat{K}^3 + b_1\, n^{\mu}n^{\nu} \hat{K}^{\alpha\beta} W_{\mu\alpha\nu\beta} \right)
 \, ,
\end{equation}
where $ \chi(M)= \int_M d^4 x \sqrt{g}E_4 - \int_M d^3 y \sqrt{h}E_{\text{bry}}$ is the Euler characteristic on manifolds with boundary. 

For ordinary (unitary) fields of spin $s=0, \, 1/2, \, 1$ for various  (Dirichlet and Robin)  conformal invariant conditions the boundary charges $b_1$ and $c$
have been calculated in \cite{Fursaev:2015wpa} while the respective values for bulk charges $a$ and $b$ were known already from previous studies.

A remarkable relation between the bulk and boundary charges was noted, namely $b_1=8b$.  Subsequently, in \cite{Solodukhin:2015eca} it was suggested that this relation
comes out naturally from the condition that the Weyl part in the integrated anomaly has a well-defined variational procedure, then the boundary term with $b_1=8b$
is precisely the Gibbons-Hawking type term. Later studies revealed a class of theories in which this relation does not hold. These typically include
some  degrees of freedom on the boundary that interact with the degrees of freedom in the bulk. Nevertheless, it is of interest to identify more precisely the class of theories in which the balance between the bulk and boundary charges takes place.  One interacting theory of this type is the CFT holographically dual to the $AdS_4$
supergravity, as was demonstrated in \cite{FarajiAstaneh:2021skf}, provided that the boundary is extended deeper in the bulk as a minimal surface. 

One of the goals of the present section is to check whether the relation holds for the (non-unitary) 4-derivative conformal field theory.
The bulk charges $a$ and $b$  for the 4-derivative conformal scalar field were computed earlier in \cite{Fradkin:1982xc,Barvinsky:1985an} using the heat kernel, and more recently in \cite{Osborn:2016bev}, where $b$ was extracted from the CFT 2-point function of the bulk energy-momentum tensor. In particular, using the results of \cite{Barvinsky:1985an} (see also \cite{deBerredo-Peixoto:2001uob}), one obtains the following values for the bulk charges $a$ and $b$,
\begin{equation}\label{eq:a_b_charges}
a=-\frac{7}{90} \, , \qquad\qquad
b=-\frac{1}{15}\, .
\end{equation}

In a CFT with a boundary, naturally appear the displacement operator $D$, which is linked to the breaking of translations in the direction normal to the boundary \cite{Herzog:2017xha}, and can be used to characterize the theory. In our case, the displacement is defined as
\begin{align}\label{eq:displacement-flat}
T^{nn}(\vec{x}_\parallel, z)|_{\partial M}
=
D(\vec{x}_\parallel)
\, .
\end{align}
The boundary charges $c$ and $b_1$ are related to the 2- and 3-point functions of $D$ through \cite{Herzog:2017kkj}
\begin{equation}\label{eq:displacement-correlators}
\langle D(x_\parallel) D(0) \rangle = \frac{15\, b_1}{2\pi^4|x_\parallel|^8}, 
\qquad
\langle D(x_\parallel) D(x'_\parallel) D(0) \rangle = \frac{35\, c}{2\pi^6|x_\parallel|^4 |x_\parallel'|^4 |x_\parallel-x_\parallel'|^4}
\, .
\end{equation}
This fact have been used in \cite{Gaikwad:2023gef} to compute these boundary charges for the a Liouville CFT and in \cite{Chalabi:2022qit} to compute $b_1$ for the $\Box^k\phi$ theory. Here, we compute $c$ and $b_1$ in the auxiliary field formalism, where we have three possible Green functions among $\phi$ and $\psi$. 

%The main difficulty is that, with boundaries, Weyl invariance in this formalism holds only on-shell, so that CFT correlators and Ward identities are valid only up to terms proportional to the equations of motion. However, $\phi$ is the same field appearing in the four order $\Box^2\phi$ CFT and as we are about to show enjoy the same 2-point function.

From the flat space limit of the Weyl invariant action \eqref{eq:weyl_invariant_auxiliary_fields_boundary}, we get the kinetic operator
\begin{align}
{\hat{D}} = 
\begin{bmatrix}
0 & \Box \\[2mm]
\Box & -1
\end{bmatrix},
\end{align}
acting on the 2-component vector $\Psi$. Its matrix valued Green's function $\hat{G}(x,x')$ is given by
\begin{align}
{\hat{D}}_x \, \hat{G}(x,x') = \hat{1}\delta^4(x-x') %\, , \qquad 
%\hat{G}(x,y)
% = 
%\begin{bmatrix}
%G_{\phi\phi}(x,x') & G_{\phi\psi}(x,x') \\
%G_{\psi\phi}(x,x') & G_{\psi\psi}(x,x') 
%\end{bmatrix}
%
\, .
\end{align}
In momentum space, we have
\begin{align}
{\hat{D}}(k)
=
\begin{bmatrix}
0 & -k^2 \\
 -k^2 & -1
\end{bmatrix}
\qquad
\Rightarrow
\qquad
{\hat{G}}(k)
=
{\hat{D}}^{-1}(k)
=
\begin{bmatrix}
\tfrac{1}{k^4} & -\tfrac{1}{k^2} \\
 -\tfrac{1}{k^2} & 0
\end{bmatrix}
\, .
\end{align}
Going back to position space, we obtain the set of 2-point functions 
\begin{subequations}
\begin{align}
\langle \phi(x)\phi(x') \rangle 
&= - \frac{1}{16\pi^2}\ln(m^2|x-x'|^2) , \label{eq:green_function_phi_phi} \\
\langle \psi(x)\phi(x') \rangle
&= \langle \phi(x)\psi(x') \rangle
= - \frac{1}{4\pi^2} \frac{1}{|x-x'|^2} , \label{eq:green_function_psi_phi} \\
\langle \psi(x)\psi(x') \rangle
&= 0  \label{eq:green_function_psi_psi}
\, ,
\end{align}
\end{subequations}
where $|x-x'|^2=|{x}_\parallel-{x}'_\parallel|^2+(z-z')^2$.
The necessary Fourier integrals are computed in the Appendix \ref{appt:fourier_Integrals}. In particular, the logarithmic 2-point function in Eq.~\eqref{eq:green_function_phi_phi} is obtained using dimensional regularization and the mass scale $m$ comes from the renormalizaiton scale $\mu$. Alternatively, it is possible to verify directly that the \eqref{eq:green_function_phi_phi} indeed satisfies $\Box^2 \langle \phi(x)\phi(x') \rangle = \delta(x,x')$, while \eqref{eq:green_function_psi_phi} satisfies $(-\Box) \langle \phi(x)\psi(x') \rangle = \delta(x,x')$. The normalizations ensures that the $\delta(x,x')$ appear with coefficient one.

In the presence of a boundary, we can use the image method to obtain the correct solution depending on boundary conditions.
We have 
\begin{subequations}\label{eq:green_functions_boundaries}
\begin{align}
&16\pi^2\langle \phi(x)\phi(x') \rangle 
=
- \ln\left(
m^2|x-x'|^2
\right) 
\pm
\ln\left(
m^2|x-x_{im}'|^2
\right)   \label{eq:green_function_phi_phi_BCs} \, ,\\
&4\pi^2\langle \phi(x)\psi(x') \rangle
=  
-\frac{1}{|x-x'|^2}
\pm
\frac{1}{|x-x_{im}'|^2}  \label{eq:green_function_psi_phi_BCs} \, ,\\
&\langle \psi(x)\psi(x') \rangle
= 0  \label{eq:green_function_psi_psi_BCs}
\, ,
\end{align}
\end{subequations}
where $x=({x}_\parallel, z)$, and $x_{im}=({x}_\parallel, -z)$ is the mirror distance. The pluses correspond to $(\phi|_{\partial M}=0,\psi|_{\partial M}=0)$ while the minuses to $(\partial_n\phi|_{\partial M}=0,\partial_n\psi|_{\partial M}=0)$.
 In the presence of a boundary, full conformal invariance is broken to the part of the conformal group leaving the boundary invariant. This allows for additional structures to appear in the corelator. In particular the 2-point function has the general form \cite{Gaikwad:2023gef}
\begin{align}
16\pi^2\langle \phi(x)\phi(x') \rangle 
=
- \ln\left(
m^2|x-x'|^2
\right) 
\pm
\ln\left(
m^2|x-x_{im}'|^2
\right)  
+
C\frac{zz'}{|x-x_{im}'|^2}
\,.
\end{align}
However, for our choice of boundary conditions $C=0$.

The next ingredient is the energy-momentum tensor.
After a lengthy computation, it is possible to show that the energy momentum tensor $T^{\mu\nu}=\tfrac{2}{\sqrt{g}}\tfrac{\delta S(\phi,\psi)}{\delta g_{\mu\nu}}$ derived from the bulk\footnote{The bulk energy-momentum tensor is not affected by the presence of the boundary. However, when computing functional derivatives with respect to the metric, to ensure a well defined variational principle and write $T_{\text{tot}}^{\mu\nu}=T^{\mu\nu} + \delta(\partial M) t^{\mu\nu}$---where $t^{\mu\nu}$ is the boundary energy momentum tensor----is crucial to verify that normal derivatives of $\delta g_{\mu\nu}$ vanish on $\partial M$. In Appendix~\ref{app:emt_auxiliary_fields} we show explicitly that this condition is satisfied for the boundary conditions $(\phi|_{\partial M}=0,\psi|_{\partial M}=0)$, and we determine the boundary terms that must be added to the action.}
 part of \eqref{eq:weyl_invariant_auxiliary_fields_boundary} satisfies
\begin{align}
\nabla_\mu T^{\mu\nu} \,\, \overset{\small{\text{on-shell}} }{=}\,\,0
\, ,
\qquad 
 T^{\mu}{}_{\mu} \,\, \overset{\small{\text{on-shell}} }{=}\,\,0
\, .
\end{align}
Moreover, its flat space limit yields
\begin{align}\label{eq:flat_space_limit_emt_axiliary_fields}
2T_{\mu\nu} =
-2 \partial_\mu \psi   \partial_\nu \phi 
+ 2 \partial_\mu \phi   \partial_\nu \psi 
&+ \tfrac{4}{3}     \partial^\alpha \phi     \partial_\alpha \partial_\mu \partial_\nu \phi  
+ 4 \partial^2 \phi    \partial_\nu \partial_\mu \phi
- \tfrac{8}{3} \partial^\alpha \partial_\mu \phi   \partial_\nu \partial_\alpha \phi
 - \delta_{\mu\nu} \Big(   
 2   (\partial^2 \phi)^2+
        \nonumber \\
&+ \tfrac{4}{3}      \partial^\alpha \phi  \partial_\alpha \partial^2 \phi    
-    \psi^2
- 2     \partial_\alpha \psi \partial^\alpha \phi
- \tfrac{2}{3}   \partial_\alpha   \partial_\rho  \phi    \partial^\alpha   \partial^\rho \phi
\Big)
\, ,
\end{align}
which coincide with the energy momentum tensor found in \cite{Stergiou:2022qqj} once we use $\psi=\Box\phi$.

 From this expression we get the normal-normal component
\begin{align}
2T^{nn} = 
-2 \partial_n \psi \partial_n \phi
+ \psi^2
+ 2\partial_a \psi \partial^a \phi
- 2 (\partial_a^2\phi)^2
- \tfrac{4}{3} \partial_a \partial_n  \phi  \partial^a  \partial_n  \phi
- \tfrac{4}{3} \partial_n\phi \partial_n \partial_a^2\phi  
- \tfrac{4}{3} \partial^b\phi \partial_b \partial_a^2\phi 
+ \tfrac{2}{3}  \partial_a \partial_b \phi \partial^a \partial^b \phi 
\,,
\end{align}
%\\ \nonumber
and using the conformally invariant boundary conditions 
\begin{align}\label{eq:conf_BC_in_flat_space}
DD: (\phi|_{\partial_M}=0, \psi|_{\partial_M}=0)\, , \qquad\quad NN: (\partial_n\phi|_{\partial_M}=0, \partial_n\psi|_{\partial_M}=0) 
\,,
\end{align}
we arrive at
\begin{align}
&(DD): \qquad 2T^{nn} 
= 
-2 \partial_n \psi \partial_n \phi
- 
\tfrac{4}{3} \partial_a \partial_n  \phi  \partial^a  \partial_n  \phi
-
 \tfrac{4}{3} \partial_n\phi \partial_n \partial_a^2\phi 
\, , \\
&(NN): \qquad 2T^{nn}
=
-2 (\partial_a^2\phi)^2
+ \psi^2 
+ 2\partial_a \psi \partial^a \phi
- \tfrac{4}{3} \partial^b\phi \partial_b \partial_a^2\phi 
+ \tfrac{2}{3}  \partial_a \partial_b \phi \partial^a \partial^b \phi 
\,,
\end{align}
whose 2- and 3-point functions can be computed via Wick's theorem and the Green functions \eqref{eq:green_functions_boundaries} to extract the boundary charges. For the displacement 2-point function and the charge $b_1$ we get
\begin{align}\label{eq:DD_correlator_DD_NN}
(DD), (NN): \qquad 
\langle D(x_\parallel) D(0) \rangle 
=
-
\frac{4}{\pi^4 |x_\parallel|^8}
\, , 
\qquad\qquad
 b_1 = -\frac{8}{15}
\, , 
\end{align}
for both pairs  of the  boundary conditions \eqref{eq:conf_BC_in_flat_space}. This result agrees with what was obtained in \cite{Gaikwad:2023gef}. We notice that these boundary conditions are imposed on different fields: 
original field $\phi$ and auxiliary fields $\psi$. The other boundary conditions that were considered in \cite{Gaikwad:2023gef} and for which value of $b_1$ differs from (\ref{eq:DD_correlator_DD_NN}) are imposed on same field (either $\phi$ or $\psi$)
while the other field remains without the boundary condition.

Comparing this with value of $b$ in Eq.~\eqref{eq:a_b_charges} we see that these results agree with the relation $b_1=8 b$ argued in \cite{Solodukhin:2015eca} on the basis of well-posedness of the variational principle.

The 3-point function and  the associated boundary charge are
\begin{align}\label{eq:DDD_correlator_DD_NN}
(DD), (NN): \qquad 
 \langle D(x_\parallel) D(x'_\parallel) D(0) \rangle 
=
-
\frac{52}{9\pi^6|x_\parallel|^4 |x_\parallel'|^4 |x_\parallel-x_\parallel'|^4}
\, ,
\qquad\qquad
 c = -\frac{104}{315}
\, .
\end{align}
In the Appendix~\ref{app:2_pt_function_displacement} we comment on the general strategy leading to Eqs~\eqref{eq:DD_correlator_DD_NN}, and \eqref{eq:DDD_correlator_DD_NN}. These results agree with \cite{Gaikwad:2023gef}---and with \cite{Chalabi:2022qit} after employing dimensional regularization---when the corresponding boundary conditions are chosen.

\section{Conclusions and outlook}\label{sect:conclusions_outlook}

In this work we developed a systematic framework for analyzing higher-derivative boundary conformal field theories in four dimensions by reformulating the biharmonic scalar theory $\Box^2$ in terms of an auxiliary second-order system. This approach greatly simplifies explicit computations while preserving the off-shell information required to study anomalies, Weyl invariance, and boundary conformal data.

In flat space, we provided a complete classification of admissible boundary conditions for the auxiliary-field formulation and established a precise correspondence with the known boundary conditions of the original fourth-order theory. It should be noticed that not all conformally invariant boundary conditions of the $\Box^2$ operator survive in the auxiliary-field framework. Although four distinct conformal boundary conditions exist at the level of the fourth-order operator, only two of them are admissible for the second-order system.
The remaining two would impose two independent conditions on the same scalar field, overconstraining the auxiliary equations and breaking ellipticity. This restriction is intrinsic to any auxiliary reformulation of higher-derivative operators and clarifies the precise domain on which the equivalence with the fourth-order theory holds. 

After diagonalizing the kinetic operator through a suitable field-space rotation, we computed the full heat kernel of $\Box^2$ exactly for a broad class of (generally non-standard) boundary conditions. This allowed us to extract the Seeley–DeWitt coefficients and to analyze the structure of the induced UV divergences, including boundary contributions associated with mixed Robin-type conditions.

On curved manifolds with boundary, we constructed the Weyl-invariant extension of the auxiliary-field action. A key result is that Weyl invariance requires the action to depend non-quadratically on the auxiliary field through the Weyl covariant combination $X=\psi-\Box\phi$. This curved-space formulation leads naturally to the identification of the conformal boundary operators and to the correct set of conformally invariant boundary conditions in the auxiliary description. The construction matches the results for the Paneitz operator with the same limitations we discussed in flat space.

Finally, making use of the auxiliary correlators and the displacement operator in boundary CFT, we computed the boundary central charges $c$ and $b_1$ by using Wick theorem. The results reproduce those obtained directly in the fourth-order formulation and are consistent with general expectations from the variational principle. This provides a nontrivial test of the auxiliary-field framework and demonstrates that it correctly captures the conformal and anomalous structure of the theory even off-shell.

There are several natural directions for future work.
The appearance of logarithmic two-point functions and the intimate relation between $\phi$ and the auxiliary field strongly suggest that the full theory can be embedded in the structure of logarithmic CFTs. This could be done by considering a near boundary expansion \cite{Hogervorst:2016itc,Bergshoeff:2011xy}. Understanding whether the energy-momentum tensor acquires a logarithmic partner---forming a logarithmic pair that mixes under dilations---and how this affects the anomaly, would offer new insight into non-unitary boundary conformal theories.
The interplay between boundary conditions and the emergence of logarithmic multiplets also deserves further investigation, especially regarding possible RG flows between different conformal boundary conditions.
Finally, our methods could be extended to higher-order GJMS operators and to the presence of spinor fields, considered in \cite{Chalabi:2022qit}. By using the result of \cite{deBerredo-Peixoto:2001uob} it would be possible to check weather the relation $b_1=8 b$ in $d=4$ between boundary and bulk charges remains true also for higher derivative spinor fields.

Overall, the auxiliary-field formalism offers a unified and computationally efficient approach to higher-derivative boundary CFTs, opening the way to a more complete understanding of conformal anomalies, boundary effects, and non-unitary structures in four dimensions.

\vspace{1.5cm}
\begin{centering}
{\textbf{Acknowledgments}}\\
\end{centering}
\vspace{.5cm}
GP wishes to thank O.~Zanusso and G.~P.~Vacca for useful discussions and for pointing out many useful references that influenced the development of this work.
The work of GP is supported by a Della Riccia foundation grant. SS would like to thank the Isaac Newton Institute for Mathematical Sciences, Cambridge, for support and hospitality during the programme {\it Quantum field theory with boundaries, impurities, and defects}, where work on this paper was undertaken. This work was supported by EPSRC grant EP/Z000580/1.

\appendix

\section{Green's formula in the presence of boundaries}\label{app:green_formula_with_boundaries}

Let $M$ be a smooth domain with boundary $\partial M$, and let $L$ be an elliptic operator acting on functions on $M$.
Suppose that $u$ solves
\begin{equation}
L u(x) = f(x), \qquad x \in M,
\end{equation}
with suitable boundary conditions on $\partial M$. The Green function $G(x,y)$ is the distributional solution of
\begin{equation}
L_x G(x,y) = \delta(x-y), \qquad x,y \in M \, .
\end{equation}
%together with the requirement that \(G\) satisfies the same boundary conditions as the problem for \(u\).
For smooth functions $v,w$ we have
\begin{equation}
\int_M \big( v Lw - w Lv \big) \, dV
= \int_{\partial M} B(v,w) \, dS,
\end{equation}
where $B(v,w)$ is the boundary antisymmetric form obtained by integrating 
by parts.

Choose $v(y) = G(x,y)$ and $w(y) = u(y)$.
We then obtain
\begin{equation}
\int_M \left( G(x,y) L_y u(y) - u(y) L_y G(x,y) \right) dV_y
= \int_{\partial M} B\!\left(G(x,\cdot),u\right) dS.
\end{equation}
Since $L_y u(y) = f(y)$ and $L_y G(x,y) = \delta(x-y)$, the bulk term gives
$
\int_M G(x,y) f(y) \, dV_y - u(x) \, .
$
This proves the general formula
\begin{equation}
u(x) = \int_M G(x,y) f(y) \, dV_y + \int_{\partial M} K(x,y) \, dS_y ,
\end{equation}
where the boundary kernel $K(x,y) \equiv -B(G(x,y),u(y))$.
The Green's formula thus consists of two contributions:
\begin{itemize}
\item[(i)] the bulk integral involving the source term $f$ and the Green function
\item[(ii)] a boundary integral, encoding the influence of the boundary data,
   where the bilinear form $B$ is determined by the structure of $L$.
\end{itemize}
%In particular, the explicit expression of \(B\) depends on the chosen operator.  
For example, for $L = \Box$ one finds
\begin{equation}
B(v,w) = v \, \partial_n w - w \, \partial_n v ,
\end{equation}
with \(\partial_n\) the normal derivative at the boundary. Thus, the Green formula becomes
\begin{equation}
u(x) = \int_M G(x,y) f(y) \, dV_y + \int_{\partial M} \left( u(y) \partial_n G(x,y) - G(x,y) \partial_n u(y)   \right) \, dS_y 
\, .
\end{equation}
We observe that, if the Green function satisfies the same boundary condition as $u$, the boundary term vanishes.

Consider now the fourth--order operator
$
L = \Box^2 .
$
For smooth functions $v,w$ one has
\begin{align}
\int_M \big( v\,\Delta^2 w - w\,\Delta^2 v \big)\,dV
= \int_{\partial M} \Big(
v\,\partial_n\Box w- \partial_n v\Box w 
 - w\,\partial_n\Box v + \partial_n w\Box v
\Big)\,dS 
\, .
\label{eq:green-bilap}
\end{align}

In general, for a \(2m\)-th order elliptic operator, the representation
formula has a similat structure: bulk integral of \(f\) against the Green
function plus a boundary integral involving \(u\) and its
normal derivatives up to order \(2m-1\).

As before, by taking $v(y)=G(x,y)$, $w(y)=u(y)$, and using 
 $\Box_y^2 u(y)=f(y)$ together with $\Box_y^2 G(x,y)=\delta(x-y)$, 
Eq.~\eqref{eq:green-bilap} yields
\begin{align}
u(x) &= \int_{\partial M} \Big[
u(y)\,\partial_{n_y}\Box_y G(x,y)
- \partial_{n_y}u(y)\,\Box_y G(x,y)
- G(x,y)\,\partial_{n_y}\Box_y u(y)
+ \partial_{n_y}G(x,y)\,\Box_y u(y) \Big] dS_y + \nonumber \\
& + \int_M G(x,y)\, f(y)
\,dV_y.
\label{eq:rep-bilap}
\end{align}
The boundary term in \eqref{eq:rep-bilap} involves 
$( u, \partial_n u, \Delta u, \partial_n \Delta u)$ and the corresponding
quantities for the Green function.  
Consider now the boundary conditions
\[
u|_{\partial M}=0, \qquad \partial_n u|_{\partial M}=0 \,.
\]
Choosing the Green function with the same conditions
$(G=0$ and $\partial_n G=0$ on $\partial M)$,
the boundary integral vanishes, and the solution depends
only on the source term $f$.
An analogous statement hold if we choose 
$
(u|_{\partial M}=0, \Box u|_{\partial M}=0) ,
$
and $(G|_{\partial M}=0, \Box_y G|_{\partial M}=0)$. It is simple to see that this is true for all the BC we considered in the main text.

Let us now consider consider the vector field
\begin{align}
u = \begin{bmatrix} u_1 \\ u_2 \end{bmatrix}
\end{align}
on a domain $M$ with boundary $\partial M$, and the off-diagonal operator
\begin{align}\label{eq:App_off_diago_op}
L =C\Box 
=
\begin{bmatrix} 0 & \Box \\ \Box & 0 \end{bmatrix} 
\,, \qquad
C = \begin{bmatrix} 0 & 1 \\ 1 & 0 \end{bmatrix}
\, .
\end{align}
We want to solve
\begin{align}
L u = f = \begin{bmatrix} f_1 \\ f_2 \end{bmatrix}, \quad x \in M
\end{align}
with some boundary conditions making $L$ self-adjoint
\begin{align}
(
\mathcal{B}_1
\Pi_{-}
u|_{\partial M}
=
0
\, ,
\quad
\mathcal{B}_2
\Pi_{+}
u|_{\partial M}
=
0
)
\, .
\end{align}
The general Green identity for $v=(v_1,v_2)^T$ is
\begin{align}
 \int_M \big( v^T L w - (L v)^T w \big) \, dV = \int_{\partial M} B(v,w) \, dS.
\end{align}
For our operator $L$ the boundary term reads explicitly
\begin{align}
\mathcal B(v,w) = v_1 \partial_n w_2 - (\partial_n v_1) w_2 + v_2 \partial_n w_1 - (\partial_n v_2) w_1,
\end{align}
as we discussed in the main text.

Consider now the $2\times2$ Green matrix
\begin{align}
G(x,y) = \begin{bmatrix} G_{11}(x,y) & G_{12}(x,y) \\[1mm] G_{21}(x,y) & G_{22}(x,y) \end{bmatrix},
\end{align}
satisfying
\begin{align}
\begin{bmatrix}
\square_x G_{21}(x,y) & \square_x G_{11}(x,y) \\
\square_x G_{22}(x,y) & \square_x G_{12}(x,y)
\end{bmatrix}
=
\begin{bmatrix}
\delta(x-y) & 0 \\
0 & \delta(x-y)
\end{bmatrix}
\, ,
\end{align}
with some boundary conditions, see below.
%with Dirichlet boundary conditions on each column:
%\begin{align}
%G_{ij}|_{\partial M} = 0, \quad i,j = 1,2.
%\end{align}
Take the first column $G_1 = \begin{bmatrix} G_{11} \\ G_{21} \end{bmatrix}$ , and set
\begin{align}
v(y)=G_1(x,y) \, ,
\qquad
w(y)=u(y)
\, .
\end{align}
With this choice the boundary term becomes
\begin{align}
B_1(G_1, w) = G_{11} \partial_n u_2 - (\partial_n G_{11}) u_2 + G_{21} \partial_n u_1 - (\partial_n G_{21}) u_1.
\end{align}
Similarly, for the second column
$
G_2 = \begin{bmatrix} G_{12} \\ G_{22} \end{bmatrix} \, ,
$
we can choose $v=G_2, \, w=u$.
The boundary term is
\begin{align}
B_2(G_2, \phi) = G_{12} \partial_n u_2 - (\partial_n G_{12}) u_2 + G_{22} \partial_n u_1 - (\partial_n G_{22}) u_1
\, .
\end{align}
These equations determine which boundary conditions the components of the Green function should satisfy, depending on those imposed on the components of $u$. For instance, as in the conformal boundary conditions discussed in the main text, if all components of $u$ satisfy the same condition, then every entry of the Green function obeys that same condition.

Then, by summing up these two results we obtain the Green representation formula
\begin{equation}
u(x) = \int_M G(x,y) f(y) \, dV_y 
+ \int_{\partial M} \Big[ (\partial_n u(y))^T C G(x,y) - u(y)^T C \, \partial_n G(x,y) \Big] \, dS_y
\, ,
\end{equation}
where $C$ is the matrix introduced in Eq.~\eqref{eq:App_off_diago_op}.% In conclusion, we have that when we impose on each column of $G$ the same BC imposed on $u$, the solution is simply
%\begin{equation}
%u(x) = \int_M G(x,y) f(y)\, dV_y
%
%\, ,
%\end{equation}
%with no additional boundary integrals.

Let us now comment on a trivial example, which nonetheless appear in the main text, and consider the operator
\begin{equation}
L = 
\begin{bmatrix}
- \Box & 0 \\[6pt]
0 & - \Box
\end{bmatrix},
\qquad 
u = 
\begin{bmatrix}
u_1 \\[4pt] u_2
\end{bmatrix}
\, .
\end{equation}
We wanto to solve
\begin{equation}
L u = f = 
\begin{bmatrix}
f_1 \\[4pt] f_2
\end{bmatrix}
\, ,
\qquad
B u|_{\partial M} = 
\begin{pmatrix}
B_1(u_1,u_2) \\[4pt] B_2(u_1,u_2)
\end{pmatrix}
=
\begin{pmatrix}
0 \\[4pt] 0
\end{pmatrix}
\end{equation}
For simplicity, we consider
\begin{equation}
((u_1 + \alpha \,\partial_n u_1)|_{\partial M} = 0 \, , \quad
(u_2 + \beta \,\partial_n u_2)|_{\partial M} = 0 )\, .
\end{equation}
Since $L$ is diagonal, the problem decouple and we have
\begin{equation}
- \Box_x \, G_{11}(x,y) = \delta(x-y), 
\qquad
- \Box_x \, G_{22}(x,y) = \delta(x-y),
\qquad
G_{12}=G_{21}=0
\, ,
\end{equation}
and
\begin{equation}
G_1(x,y) = 
\begin{pmatrix}
G_{11}(x,y) \\[4pt] 0
\end{pmatrix},
\qquad
G_2(x,y) = 
\begin{pmatrix}
0 \\[4pt] G_{22}(x,y)
\end{pmatrix}
\, .
\end{equation}
Proceeding as before, we set
$
v = G_i(x,\cdot)
$
and
$
w = u
$ with $i=1,2$. Thus, we have
\begin{equation}
\int_M G_{ii}(x,y) f_i(y)\, dV_y 
= u_i(x) - \int_{\partial M} 
\bigl( G_{ii}\,\partial_n u_i - u_i \,\partial_n G_{ii} \bigr) dS_y
\, ,
\end{equation}
or, in matrix notation,
\begin{equation}
u(x) = \int_M G(x,y) f(y) \, dV_y 
+ \int_{\partial M} \Big[ (\partial_n u(y))^T G(x,y) - u(y)^T  \partial_n G(x,y) \Big] \, dS_y
\, .
\end{equation}
Notice that for $L=\sigma_3 \Box$, i.e., with opposite signs in the box operators, the Green matrix must be multiplied by $\sigma_3$, similarly to the previous example.

\section{Fourier Integrals}\label{appt:fourier_Integrals}

In this appendix, we give some details on the computations leading Eq.s~\eqref{eq:green_function_phi_phi}, \eqref{eq:green_function_psi_phi} and \eqref{eq:green_function_psi_psi}. The necessary Fourier integrals have the standard form
\begin{align}\label{eq:fourier_transform_formula}
I_\alpha(x) 
=
\int \frac{d^d k}{(2\pi)^d} \frac{e^{i k \cdot x}}{(k^2)^\alpha} 
 \, .
\end{align}
To compute this integral, we start with the Schwinger representation
\begin{align}
\frac{1}{A^\alpha} = \frac{1}{\Gamma(\alpha)} \int_0^\infty ds \, s^{\alpha-1} e^{-s A}
\, .
\end{align}
This is easily proved from the definition of the $\Gamma$ function,
${\small{
\Gamma(\alpha)
=
\int_0^\infty ds s^{\alpha-1}  e^{-s} 
=
A^\alpha \int_0^\infty dt \, t^{\alpha-1} e^{-At} }}
\, ,
$
where we used the change of variables $s=At$.
Then, for $A=k^2$, we have
\begin{align}
I_\alpha(x)
= \frac{1}{\Gamma(\alpha)} \int_0^\infty ds \, s^{\alpha-1} \int \frac{d^d k}{(2\pi)^d} \, e^{-s k^2 + i k \cdot x}
\, .
\end{align}
The integral over $k$ is Gaussian
${\small{
\int d^d k \, e^{-s k^2 + i k \cdot x} = \left(\frac{\pi}{s}\right)^{d/2} e^{- \frac{x^2}{4s}}}}
\, .
$
 Thus, including the $(2\pi)^{-d}$ factor
\begin{align}
I_\alpha(x) = \frac{1}{\Gamma(\alpha)} \frac{1}{(4\pi)^{d/2}} \int_0^\infty ds \, s^{\alpha-1-d/2} e^{- \frac{x^2}{4s}}.
\end{align}
Performing the change of variable \(t = \frac{x^2}{4s} \Rightarrow s = \frac{x^2}{4 t}, \, ds = - \frac{x^2}{4 t^2} dt\), the integral becomes
\begin{align}
I_\alpha(x) = \frac{1}{\Gamma(\alpha)} \frac{1}{(4\pi)^{d/2}} \left( \frac{x^2}{4} \right)^{\alpha - d/2} \int_0^\infty dt \, t^{d/2 - \alpha - 1} e^{-t}.
\end{align}
The integral over $t$ is a Gamma function
${\small{
\int_0^\infty dt \, t^{d/2 - \alpha - 1} e^{-t} = \Gamma\left(\frac{d}{2} - \alpha\right) }}
\, .
$
Thus, we arrive at the general formula 
\begin{align}
I_\alpha(x) 
=
\int \frac{d^d k}{(2\pi)^d} \, \frac{e^{i k \cdot x}}{(k^2)^\alpha} 
= 
\frac{1}{(4\pi)^{d/2}} \frac{\Gamma\left(\frac{d}{2}-\alpha\right)}{\Gamma(\alpha)} 
\left( \frac{x^2}{4} \right)^{\alpha - d/2}.
\end{align}
This is the key formula from which the two cases of interest are derived. The first case corresponds to the choices $\alpha=1$ and $d=4$. Thus, we immediately obtain
\begin{align}
I_1(x)
=
\frac{1}{4\pi^2}\frac{1}{x^2}
\, .
\end{align}
The second corresponds to $\alpha=2$ and $d=4$, so that $I_2(x)$ is logarithmic divergent. We adopt dimensional regularization and set $d=4-\epsilon$. Thus we have
\begin{align}
I_2(x) 
=
(\mu^2)^{\tfrac{\epsilon}{2}}
\int \frac{d^{4-\epsilon} k}{(2\pi)^{2-\epsilon/2}} \, \frac{e^{i k \cdot x}}{(k^2)^\alpha} 
=
 \frac{        (\mu^2)^{\tfrac{\epsilon}{2}}    }{(4\pi)^{2-\epsilon/2}}  \frac{    \Gamma      \left(\tfrac{\epsilon}{2}  \right)    }{\Gamma(2)} 
 \left( \tfrac{x^2}{4} \right)^{     \tfrac{\epsilon}{2}   }.
\end{align}
By using the familiar expansions
\begin{align}
\Gamma      \left(\tfrac{\epsilon}{2}  \right)  = -\tfrac{2}{\epsilon} - \gamma + O(\epsilon) 
\, ,  
\quad\quad 
(4\pi)^{-\epsilon/2}  = 1 - \tfrac{\epsilon}{2} \ln(4\pi)  + O(\epsilon) 
\, ,  
\quad\quad
 \left( \tfrac{\mu^2x^2}{4} \right)^{     \tfrac{\epsilon}{2} }   = 1 + \tfrac{\epsilon}{2} \ln \tfrac{\mu^2x^2}{4}  + O(\epsilon) 
 \, ,
\end{align}
we get
\begin{align}
I_2(x) 
=
\frac{1}{16\pi^2}
\left(
-\frac{2}{\epsilon}
-
\ln \left(\tfrac{    {\tilde{\mu}}^2x^2}{4}\right)
\right)
\end{align}
where, as customary in the $\overline{MS}$-scheme, we redefined $ {\tilde{\mu}}^2=\mu^2 \tfrac{e^\gamma}{4\pi}$ to absorb the constants. Thus, after renaming ${\tilde{\mu}} \sim m$ and renormalization, these correspond to the results reported in the main text.

\mycomment{
More explictly, we have
\begin{align}
\Gamma^{D(N)}_{eff,\Box}
=
-\frac{1}{2}
V_{d-1}
\int_{\epsilon^2}^{\infty}  
\frac{ds}{s}
\int_{\mathbb{R}^+} dz 
\frac{1}{(4\pi s)^{d/2}}
\left[
\left(
1
-
e^{-\frac{z^2}{s}}
\right)
\left( 
1 + \frac{sm^2}{2}
\right)
+
\left(
1
+
e^{-\frac{z^2}{s}}
\right)
\left( 
1 - \frac{sm^2}{2}
\right)
\right]
\, ,
\end{align}

}

%%%%%%%%%%%%%%%%%%%%%%%%%%%%%%%%%%%%%%%%%%%%%%%%%%%%%%%%%%%%%%%%%%%%%%%%%%%%%%%%%%%%%%%%%%%%%%%%%%%%%%%%%%%%%%%%%%%%%%%%%%%%%%%%%%%%%%%%%%%%%%%%%%%%%%%%%%%%%%%%%%%%%%%%%%%%%%%%%%%%%%%%%%%%%%%%%%%%%%%%%%%%%%%%%%%%%%%%%%%%%%%%%%%%%%%%%%%%%%%%%%%%%%%%%%%%%%%%%%%%%%%%%%%%%%%%%%%%%%%%%%%%%%%%%%%%%%%%%%%%%%%%%%%%%%%%%%%%%%%%%%%%%%%%%%%%%%%%%%%%%%%%%%%%%%%%%%%%%%%%%%%%%%%%%%%%%%%%%%%%%%%%%%%%%%%%%%%%%%%%%%%%%%%%%%%%%%%%%%%%%%%%%%%%%%%%%%%%%%%%%%%%%%%%%%%%%%%%%%%%%%%%%%%%%%%%%%%%%%%%%%%%%%%%%%%%%%%%%%%%%%%%%%%%%%%%%%%%%%%%%%%%%%%%%%%%%%%%%%%%%%%%%%%%%%%%%%%%%%%%%%%%%%%%%%%%%%%%%%%%%%%%%%%%%%%%%%%%%%%%%%%%%%%%%%%%%%%%%%%%%%%%%%%%%%%%%%%%%%%%%%%%%%%%%%%%%%%%%%%%%%%%%%%%%%%%%%%%%%%%%%%%%%%%%%%%%%%%%%%%%%%%%%%%%%%%%%%%%%%%%%%%%%%%%%%%%%%%%%%%%%%%%%%%%%%%%%%%%%%%%%%%%%%%%%%%%%%%%%%%%%%%%%%%%%%%%%%%%

\section{Energy momentum tensor in the auxiliary field formalism}\label{app:emt_auxiliary_fields}
We want to compute the metric variation $\delta=\delta_g$ of the action
\begin{align}\label{eq:initial_action}
 S(\phi,\psi)
 &=
\int_M d^4  x \sqrt{ g} 
\left(
  \phi \Box \psi
  + \psi \Box \phi
  - \psi^2
    +
  \phi
  \nabla_\mu
  \left(V^{\mu\nu}
\partial_\nu
\phi
\right)
\right)
+
\tfrac{2}{3}
\int_{\partial M} \hspace{-.1cm} d^3 y  \sqrt{h}
\phi K X \\ \nonumber
&\equiv
\int_M \hspace{-.1cm}  \sqrt{g} \mathcal{L}_V
+
\int_{\partial M} \hspace{-.2cm}  \sqrt{h}\mathcal{L}_B
\, ,
\end{align}
induced by the variation $\delta g_{\alpha\beta}$. Ultimately, we will do so for the conformal boundary conditions
\begin{align}\label{eq:conf_dirchlet}
\left(\phi|_{\partial M}=0, \quad\left.\left(\psi - \tfrac{2}{3}K\partial_n\phi \right)\right|_{\partial M}=0\right)
\, .
\end{align}
However, we keep the computation as general as possible.
Schematically, we have
\begin{align}
\delta S(\phi,\psi)
=
\int_M (\delta \sqrt{g}) \mathcal{L}_V
+
\int_{\partial M} \hspace{-.1cm}  (\delta \sqrt{h}) \mathcal{L}_B
+
\int_M \hspace{-.1cm}  \sqrt{g} (\delta\mathcal{L}_V)
+
\int_{\partial M} \hspace{-.2cm}  \sqrt{h} (\delta \mathcal{L}_B)
\, .
\end{align}
As usual, $\delta\sqrt{g} = \tfrac{1}{2} \sqrt{g} g^{\mu\nu}\delta g_{\mu\nu}$ and similarly for $\delta \sqrt{h}$. Notice that $h^{\mu\nu}\delta h_{\mu\nu}=h^{\mu\nu}\delta g_{\mu\nu}$.
We focus on the last two terms, as varying the curvature tensors produces normal derivatives of the metric variation. We start with $\delta\mathcal{L}_V$.
Under a metric variation the Christoffel symbols yield
 \begin{align}\label{eq:metric_variation_christoffel}
\delta \Gamma^{\lambda}{}_{\mu \nu} 
=
 \tfrac{1}{2} g^{\lambda \rho} 
\left( 
\nabla_{\mu} \delta g_{\rho \nu} 
+ \nabla_{\nu} \delta g_{\mu\rho} 
- \nabla_{\rho}\delta g_{\mu\nu}
\right)
\, .
\end{align}
For the Laplacian of a scalar field we have 
\begin{align}
\delta(g^{\mu\nu}\nabla_\mu\partial_\nu \phi)
=
\delta g^{\mu\nu} \nabla_\mu \partial_\nu \phi
+
g^{\mu\nu}  \delta\Gamma^{\lambda}{}_{\mu \nu} \partial_\lambda \phi
=
\delta g^{\mu\nu} \nabla_\mu \partial_\nu \phi
-
g^{\mu\nu}
\left(
\tfrac{1}{2} \nabla_\rho \delta g_{\mu\nu} 
-
\nabla_\mu \delta g_{\nu\rho} 
\right)
\partial^\rho \phi
\, .
\end{align}
Using this equation and integrating by parts we get
\begin{align}
\int_M \sqrt{g} \delta (\psi \Box \phi)
&=
\int_M \sqrt{g} 
\left(
\partial^\mu\psi \partial^\nu \phi
-
\tfrac{1}{2} g^{\mu\nu}\nabla_\rho(\psi \partial^\rho \phi)
\right)\delta g_{\mu\nu} \, + \\ \nonumber
&+ \tfrac{1}{2}  \int_{\partial M} \sqrt{h} 
\left(
g^{\mu\nu} \psi \partial_n\phi
-
n^\mu \psi \partial^\nu \phi
-
n^\nu \psi \partial^\mu \phi
\right)\delta g_{\mu\nu}
\, .
\end{align}
An analogous result holds for $\phi\Box\psi$ by exchanging $\psi$ with $\phi$. Summing these results we get the boundary term
\begin{align}
 \tfrac{1}{2}  \int_{\partial M} \sqrt{h} 
 \left[
\psi
\left(
g^{\mu\nu}  \partial_n\phi
-
n^\mu  \partial^\nu \phi
-
n^\nu \partial^\mu \phi
\right)
+
\phi
\left(
g^{\mu\nu}  \partial_n\psi
-
n^\mu  \partial^\nu \psi
-
n^\nu \partial^\mu \psi
\right)
\right]
\delta g_{\mu\nu}
\, ,
\end{align} 
where no derivatives of $\delta g$ appear.
For the curvature tensors we have
\begin{align}
&\delta R^{\lambda}{}_{\mu \sigma \nu} 
= \nabla_{\sigma} \delta \Gamma^{\lambda}{}_{\mu\nu}
- \nabla_{\nu} \delta \Gamma^{\lambda}{}_{\mu\sigma}
\, , \\
&\delta R_{\mu\nu} 
= \nabla_{\lambda} \delta \Gamma^{\lambda}{}_{\mu\nu}
- \nabla_{\nu} \delta \Gamma^{\lambda}{}_{\mu\lambda} 
= \tfrac{1}{2} \left( 
\nabla^{\lambda}  \nabla_{\mu} \delta g_{\lambda \nu}
+ \nabla^{\lambda}  \nabla_{\nu} \delta g_{\mu \lambda} 
- g^{\lambda \rho} \nabla_{\mu} \nabla_{\nu} \delta g_{\lambda \rho} 
- \Box \delta g_{\mu\nu} 
\right)
\, , \\
&\delta R 
= 
- R^{\mu\nu} \delta g_{\mu\nu} 
+ \nabla^{\mu} \left( 
\nabla^{\nu} \delta g_{\mu\nu} 
- g^{\lambda \rho} \nabla_{\mu} \delta g_{\lambda \rho} 
\right)
\, .
\end{align}
Combining these results, we can easily compute the variation of $V^{\mu\nu}=2R^{\mu\nu} -\tfrac{2}{3} g^{\mu\nu}R$. We get 
\begin{align}\label{eq:metric_variation_V}
\delta V^{\mu\nu}
&= g^{\mu\alpha} g^{\nu\beta} \Big(
    \nabla^\lambda  \nabla_\alpha  \delta g_{\lambda\beta}
   + \nabla^\lambda  \nabla_\beta \delta g_{\lambda\alpha}
   - g^{\lambda\rho}\nabla_\alpha \nabla_\beta \delta g_{\lambda\rho}
   - \Box \delta g_{\alpha\beta}
   \Big) 
\\ \nonumber
&\quad
- 2 \big( R^{\mu\alpha}  \delta g^{\nu}{}_{\alpha}
        +  R^{\nu\alpha} \delta g^{\mu}{}_{\alpha}  \big)
+\tfrac{2}{3} \delta g^{\mu\nu} R
- \tfrac{2}{3} g^{\mu\nu}
  \big( \nabla^\alpha \nabla^\beta \delta g_{\alpha\beta}
       - g^{\alpha\beta}\Box  \delta g_{\alpha\beta}
       - R^{\alpha\beta} \delta g_{\alpha\beta} \big)
\, ,
\end{align}
where $\delta g^{\mu}{}_{\alpha}=g^{\mu\beta}\delta g_{\beta\alpha}$.
For the normal vector we get
\begin{align}\label{eq:metric_variation_normal}
\delta n_{\mu} 
= \tfrac{1}{2} n_{\mu} n^{\nu} n^{\lambda} \delta g_{\nu\lambda} \equiv a \, n_\mu
\, ,
\qquad 
\delta n^{\mu} 
=a \, n^\mu - g^{\mu\alpha} n^\beta \delta g_{\alpha\beta} 
\, ,
\end{align}
obtained from $\delta(n_\mu n^\mu)=0$ and $\delta(e^\mu_a n_\mu)=0$. In the last variation we keep the boundary fixed, meaning that $\delta e^\mu_a =0$. From these equations we obtain the variation of the projector  $h^{\alpha\beta}=g^{\alpha\beta} - n^\alpha n^\beta$ as
\begin{align}\label{eq:variation_projector}
\delta h_{\mu}{}^{\alpha}= - 2 a n_\mu n^\alpha + n_\nu g^{\alpha\rho} \delta g_{\rho\gamma} n^\gamma
\, .
\end{align}
For the extrinsic curvature $K_{\mu\nu}=h_{\mu}{}^{\alpha} h_{\nu}{}^{\beta} \nabla_\alpha n_\beta$ the variation yields
\begin{align}
\delta K_{\mu\nu}
=
(\delta h_{\mu}{}^{\alpha}) h_{\nu}{}^{\beta} \nabla_\alpha n_\beta
+
h_{\mu}{}^{\alpha} (\delta h_{\nu}{}^{\beta}) \nabla_\alpha n_\beta
+
h_{\mu}{}^{\alpha} h_{\nu}{}^{\beta}\left( \nabla_\alpha \delta n_\beta - n_\lambda \delta \Gamma^\lambda_{\alpha\beta}\right)
\, .
\end{align}
By using the decomposition
\begin{align}\label{eq:decomposing_nabla_n}
\nabla_\mu n_\nu 
=
K_{\mu\nu} +  n_\mu a_\nu
=
 K_{\mu\nu} +  n_\mu n^\lambda \nabla_\lambda n_\nu
 \, ,
\end{align}
Eq.~\eqref{eq:variation_projector} and the transversality of $K_{\mu\nu}$ and $a_\nu$, the first two terms combine into
\begin{align}
(\delta h_{\mu}{}^{\alpha}) h_{\nu}{}^{\beta} \nabla_\alpha n_\beta
+
h_{\mu}{}^{\alpha} (\delta h_{\nu}{}^{\beta}) \nabla_\alpha n_\beta
=
\delta g_{\rho\gamma} n^\gamma \left( n_\mu K^{\rho}{}_\nu + n_\nu K^{\rho}{}_\mu  \right)
\, .
\end{align}
Including the contributions of the last two terms and using Eq.s~\eqref{eq:metric_variation_christoffel} and \eqref{eq:metric_variation_normal}, we arrive at
\begin{align}\label{eq:metric_variation_K}
\delta K_{\mu\nu}
=
\delta g_{\alpha\beta} n^\beta \left( n_\mu K^{\alpha}{}_\nu + n_\nu K^{\alpha}{}_\mu  \right)
+
\tfrac{1}{2} K_{\mu\nu}  n^{\alpha} n^{\beta} \delta g_{\alpha\beta} 
-
\tfrac{1}{2} 
h_{\mu}{}^{\alpha} h_{\nu}{}^{\beta} 
n^\lambda
\left( 
\nabla_{\alpha} \delta g_{\lambda \beta} 
+\nabla_{\beta} \delta g_{\lambda \alpha} 
- \nabla_{\lambda}\delta g_{\alpha\beta}
\right)
\, .
\end{align}
For the trace of the extrinsic curvature $K=g^{\mu\nu}K_{\mu\nu}$, we have
%\begin{align}
%c_{\mu} 
%= n_{\mu} n^{\nu} n^{\lambda} \delta g_{\nu\lambda} - \delta g_{\mu\nu} n^{\nu}
%= - h_{\mu}{}^{\lambda} \delta g_{\lambda \nu} n^{\nu}
%\end{align}
\begin{align}\label{eq:metric_variation_trace_K}
\delta K 
=
-
K^{\mu\nu} \delta g_{\mu\nu}
+
g^{\mu\nu} \delta K_{\mu\nu}
=
 -
K^{\mu\nu} \delta g_{\mu\nu} 
+ 
\tfrac{1}{2} K\, n^{\alpha} n^{\beta} \delta g_{\alpha\beta} 
+ 
 h^{\alpha\beta} n^\lambda
\left( 
\tfrac{1}{2} \nabla_\lambda \delta g_{\alpha\beta}
- \nabla_\alpha \delta g_{\beta\lambda}
\right)%
\, ,
\end{align}
where we used $n_\mu K^{\mu\nu}=0$ and $h^{\alpha\mu}h_{\mu}{}^{\beta}=h^{\alpha\beta}$.
Deriving these equations it is also useful to remember that $g^{\alpha\beta}\delta g_{\alpha\beta}=- g_{\alpha\beta}\delta g^{\alpha\beta}$.
Notice that assuming $\delta g_{\mu\nu}|_{\partial M}=0$ as usually done for the variational principle, we have $\delta K = \tfrac{1}{2} n^\mu h^{\rho\lambda} \nabla_\mu \delta g_{\rho\lambda}$.

To avoid third order derivatives on $\delta g$, we integrate by part the last contribution in $\mathcal{L}_V$ and compute the variation obtaining
\begin{align}\label{eq:curvature_dep_terms}
- \int_M d^4  x \sqrt{ g} \,
 \delta V^{\mu\nu}
 \partial_\mu \phi
\partial_\nu \phi
+
\int_{\partial M} \hspace{-.1cm} d^3 y  \sqrt{h} \,
\phi \delta V^{n\nu} \partial_\nu \phi
\, .
\end{align}
Inserting Eq.~\eqref{eq:metric_variation_V} in the volumetric integral and keeping only contributions displaying derivatives of the variation, of the metric we obtain
\begin{align}\label{eq:variation_V_derivatives_deltag}
\int_M d^4  x \sqrt{ g} \,
\left[
\tfrac{2}{3} (\partial_\mu \phi)^2
\left(
 \nabla^\beta \nabla^\alpha \delta g_{\alpha\beta}
-
\Box \delta g 
\right)
+
\partial^\alpha \phi \partial^\beta \phi
\left(
\Box  \delta g_{\alpha\beta}
+
\nabla_\alpha \nabla_\beta \delta g
-
2\nabla_\alpha \nabla_\rho \delta g^{\rho}_{\beta}
\right)
\right]
\, ,
\end{align}
where $\delta g = g^{\alpha\beta}  \delta g_{\alpha\beta}$. Integrating by parts the first two terms we obtain the following boundary terms 
\begin{align}\label{eq:normal_derivatives_1}
\tfrac{2}{3} \int_M d^4  x \sqrt{ g} \,
(\partial_\mu \phi)^2
\left(
 \nabla^\beta \nabla^\alpha \delta g_{\alpha\beta}
-
\Box \delta g 
\right)
&=
-\tfrac{2}{3} \int_{\partial M} d^3  x \sqrt{h} \,
(\partial_\mu \phi)^2
\left(
\partial_n \delta g
-
n^\beta \nabla^\alpha \delta g_{\alpha\beta}
\right)
+ \dots \\ \nonumber
&=
-\tfrac{2}{3} \int_{\partial M} d^3  x \sqrt{h} \,
(\partial_\mu \phi)^2
\left(
h^{\alpha\beta} n^\lambda
\nabla_\lambda \delta g_{\alpha\beta}
-
n^\beta D^\alpha \delta g_{\alpha\beta}
\right)
+ \dots
\, ,
\end{align}
where the dots stand for contributions $\propto \delta g_{\alpha\beta}, D_\rho g_{\alpha\beta}$ and we decomposed in tangential and normal components $\nabla^\alpha \delta g_{\alpha\beta}= D^\alpha \delta g_{\alpha\beta} + n^\alpha \nabla_n  \delta g_{\alpha\beta}$. 
Similarly the last three contributions of Eq.~\eqref{eq:variation_V_derivatives_deltag} yield
\begin{align}\label{eq:normal_derivatives_2}
&\int_M d^4  x \sqrt{ g} \,
\partial^\alpha \phi \partial^\beta \phi
\left(
\Box  \delta g_{\alpha\beta}
+
\nabla_\alpha \nabla_\beta \delta g
-
2\nabla_\alpha \nabla_\rho \delta g^{\rho}_{\beta}
\right)
=
\int_{\partial M} d^3  x \sqrt{h} \,
 \left(
\partial^\alpha \phi \partial^\beta \phi \,
\nabla_n \delta g_{\alpha\beta} + \right. \\ \nonumber
&\left.
+
\partial_n \phi \partial^\mu \phi \, \nabla_\mu \delta g
-
2\partial_n \phi \partial^\mu \phi  \, \nabla_\rho \delta g^{\rho}_\mu
\right)
+ \dots 
=
\int_{\partial M} d^3  x \sqrt{h}
 \left(
 (\partial_n \phi)^2 h^{\alpha\beta} n^\lambda
\nabla_\lambda \delta g_{\alpha\beta}
 +
 \partial^\mu \phi \partial^\nu \phi \,
 h_{\mu}{}^{\beta} h_{\nu}{}^{\gamma}n^\lambda
   \nabla_\lambda \delta g_{\beta\gamma}
 \right)
+ \dots 
\, ,
\end{align} 
Eq.~\eqref{eq:metric_variation_trace_K} and \eqref{eq:metric_variation_K} shows that to cancel the normal derivatives in \eqref{eq:normal_derivatives_1}  and \eqref{eq:normal_derivatives_2}, we can add the surface integral
\begin{align}\label{eq:boundary_term_for_metric_variation}
B_{K}=&\int_{\partial M} d^3  x \sqrt{h} \,
\left(
\tfrac{4}{3}K \partial_\mu \phi \partial^\mu \phi
-
2 K \partial_n \phi  \partial_n \phi
-
2 K_{\mu\nu}  \partial^\mu \phi \partial^\nu \phi 
\right) \\ \nonumber
&=
-2
\int_{\partial M} d^3  x \sqrt{h} \,
\left(
\tfrac{1}{3}K (\partial_n \phi)^2
+
 \tfrac{1}{3} K (\partial_a \phi)^2
+
 \tilde{K}_{ab}  \partial^a \phi \partial^b \phi 
\right)
\, .
\end{align}
This boundary integral is necessary to have a consistent metric variation but it breaks Weyl invariance. Let us now consider the conformal BCs \eqref{eq:conf_dirchlet}. Under these conditions, the boundary term of Eq.~\eqref{eq:boundary_term_for_metric_variation} transform under Weyl as
\begin{align}\label{eq:Weyl_variation_B_k}
\delta^W_\sigma B_{K}
=
\int_{\partial M} d^3  x \sqrt{h} \,
\partial_n \sigma (\partial_n \phi)^2
\, , 
\end{align}
which can be killed by $B_\psi =  \int_{\partial M} d^3  x \sqrt{h} \psi \partial_n \phi$. Another great simplifiction offered by \eqref{eq:conf_dirchlet} is that the boundary integrals in Eq.s~\eqref{eq:initial_action} and \eqref{eq:curvature_dep_terms} vanish. In conclusion, we have that the action
\begin{align}
S_{\text{tot}}
=
S(\phi,\psi)
+
B_K
+
B_\psi
\end{align}
is Weyl invariant when the fields satisfy the BCs \eqref{eq:conf_dirchlet} and its metric variation can be written as
\begin{align}
\delta_g S_{\text{tot}}
=
-\tfrac{1}{2}\int_M  \sqrt{g}  T^{\mu\nu} \delta g_{\mu\nu}
-
\tfrac{1}{2} \int_{\partial M} \hspace{-.1cm}  \sqrt{h} t^{ab} \delta h_{ab}
\, .
\end{align}
 Then, the energy momentum tensor can be written as
\begin{align}
T_{\text{tot}}^{\mu\nu}=\tfrac{-2}{\sqrt{g}}\tfrac{\delta S(\phi,\psi)}{\delta g_{\mu\nu}}
=
T^{\mu\nu} 
+
\delta(\partial M) t^{\mu\nu}
\, ,
\end{align}
where  $T^{\mu\nu} $ is reported in Eq.~\eqref{eq:flat_space_limit_emt_axiliary_fields} of the main text, and $\delta(\partial M)$ is a delta function with support on the boundary. $t^{\mu\nu}$ is the boundary energy-momentum tensor and its expression could be found with the equation presented in this appendix. However, for us its important properties is that $t^{\mu\nu}$ is a tangential tensor field, i.e., $t^{\mu\nu} =t^{ab} e^{\mu\vphantom{\nu}}_{a\vphantom{b}} e^{\nu\vphantom{\mu}}_{b\vphantom{a}}$, so that its normal components vanish and we have
\begin{align}
T_{\text{tot}}^{nn}
=
T^{nn} 
\, .
\end{align}

\section{Two-point function of the displacement operator using Wick theorem}\label{app:2_pt_function_displacement}
In this Appendix we briefly outline the computation of correlators of the displacement operator. 
For simplicity, we focus on the two-point function with $(DD)$ boundary conditions. 
The computation of three-point functions or correlators with different boundary conditions follows the same steps, although the algebra becomes more involved.

We aim to compute the two-point function
\begin{align}
(DD): \qquad \langle D(x_\parallel) D(0)   \rangle = \langle T_1^{nn} T_2^{nn} \rangle|_{\partial M}
\, ,
\end{align}
where $1 \equiv (x_\parallel, z)$, $2 \equiv (x'_\parallel, z)$, and
\begin{align}
2T^{(DD)}_{nn} 
= 
-2 \partial_n \psi \partial_n \phi
- 
\frac{4}{3} \partial_a \partial_n  \phi  \partial^a  \partial_n  \phi
-
 \frac{4}{3} \partial_n\phi \partial_n \partial_a^2\phi 
\equiv
A(x)
+
B(x)
+
C(x) 
\, ,
\end{align}
where $\partial_n \equiv \partial_z$.
Excluding the disconnected vacuum bubble, the Wick theorem gives
\begin{align}\label{eq:wick_theo}
\langle D_1 D_2 \rangle_{\mathrm{conn}}
= \sum_{X,Y \in \{A,B,C\}} \langle X_1 Y_2 \rangle_{\mathrm{conn}} \,.
\end{align}
The relevant propagators are
\begin{align}
&16\pi^2\langle \phi(x)\phi(x') \rangle 
=
- \ln\left(
m^2\left(r^2 + (z-z') \right)
\right) 
\pm
\ln\left(
m^2\left(r^2 + (z+z') \right)
\right)  \\
&4\pi^2\langle \phi(x)\psi(x') \rangle
=  
-\frac{1}{r^2 + (z-z')}
\pm
\frac{1}{r^2 + (z+z')} \\
&\langle \psi(x)\psi(x') \rangle
=  
0
\,,
\end{align}
where $r^2=|x_\parallel-x'_\parallel|^2$. 
The normal derivatives evaluated on the boundary are
\begin{align}\label{eq:normal_deriv_propgators}
\partial_z\partial_{z'} \langle \phi(x)\phi(x') \rangle|_{\partial M}
=
\frac{1}{4\pi^2 r^2}  \, , \qquad\qquad 
\partial_z\partial_{z'} \langle \phi(x)\psi(x') \rangle|_{\partial M}
=  
-\frac{1}{\pi^2 r^4}
\,.
\end{align}
We also use the identities
\begin{align}\label{eq:derivatives_r}
\partial_a \frac{1}{r^p} = - p \frac{ \left(x_\parallel - x'_\parallel \right)_a}{r^{p + 2}} \, , \qquad\qquad \partial^2_a \frac{1}{r^p} = \frac{p (p + 2 -d)}{r^{p + 2}} \, ,
\end{align}
where $\partial_a \equiv \partial_{x^a_\parallel}$. Taking the derivative  $\partial_{a'} \equiv \partial_{{x'}^a_\parallel}$ the first relation acquires a minus sign.
From these simple relations it is straightforward to compute all required correlators. 
For instance, in  $d=3$, we have
\begin{align}
\partial_a\partial_{b'}\partial_z\partial_{z'} \langle \phi(x)\phi(x') \rangle|_{\partial M} = \frac{2}{\pi^2}\left(\tfrac{\delta_{ab'}}{r^4} - \tfrac{ \left(x_\parallel - x'_\parallel \right)_a  \left(x_\parallel - x'_\parallel \right)_{b'}}{r^6}    \right)  \, , \qquad\qquad 
\partial^2_a\partial_z\partial_{z'} \langle \phi(x)\psi(x') \rangle|_{\partial M} = - \frac{12}{\pi^2 r^6}
\,.
\end{align}
Using Eq.s~\eqref{eq:normal_deriv_propgators} and \eqref{eq:derivatives_r} into \eqref{eq:wick_theo}, in the limit $x'_\parallel \to 0$, we arrive at the result
\begin{align}
(DD): \qquad \langle D(x_\parallel) D(0) \rangle 
=
-
\frac{4}{\pi^4 |x_\parallel|^8}
\, ,
\end{align}
reported in the main text. The value of the boundary charge can be read off Eq.s~\eqref{eq:displacement-correlators}.

\end{document}